\documentclass[twocolumn,prl,aps,showpacs,showkeys,amsmath,amssymb,superscriptaddress,final,reprint,floatfix,longbibliography]{revtex4-1}
\usepackage{color}
\usepackage{amsmath}
\usepackage{amssymb}
\usepackage{stmaryrd}
\usepackage{graphicx}
\usepackage[bookmarks=false,colorlinks=true, citecolor=blue, urlcolor=blue ]{hyperref}
\usepackage{pdfpages} 
\usepackage{tikz} 
\makeatletter 
\AtBeginDocument{\let\LS@rot\@undefined} 
\makeatother 

\usepackage{newtxtext}
\usepackage[libertine]{newtxmath}

\usepackage[all]{hypcap}

\newcommand{\ket}[1]{\left|#1\right>}

\newcommand{\blb}[1]{\left[ #1 \right]}
   \definecolor{mybrown_fig}{rgb}{0.85,0.325,0.098}
   \definecolor{myblue_fig}{rgb}{0,0.45,0.74}

\begin{document}
\title{Experimental measurement of the quantum geometric tensor \\ using coupled qubits in diamond}
\author{Min~Yu}
\thanks{These authors contribute equally.}
\author{Pengcheng~Yang}
 \thanks{These authors contribute equally.}
\author{Musang~Gong}
\affiliation{School of Physics, Huazhong University of Science and Technology, Wuhan 430074, China}
\affiliation{International Joint Laboratory on Quantum Sensing and Quantum Metrology, Huazhong University of Science and Technology, Wuhan 430074, China}
\author{Qingyun Cao}
\affiliation{Institut f\"{u}r Quantenoptik $\&$ IQST, Albert-Einstein Allee 11, Universit\"{a}t Ulm, D-89081, Germany}
\affiliation{School of Physics, Huazhong University of Science and Technology, Wuhan 430074, China}
\affiliation{International Joint Laboratory on Quantum Sensing and Quantum Metrology, Huazhong University of Science and Technology, Wuhan 430074, China}
\author{Qiuyu Lu}
\author{Haibin Liu}
\affiliation{School of Physics, Huazhong University of Science and Technology, Wuhan 430074, China}
\affiliation{International Joint Laboratory on Quantum Sensing and Quantum Metrology, Huazhong University of Science and Technology, Wuhan 430074, China}
\author{Shaoliang~Zhang}
\email{shaoliang@hust.edu.cn}
\affiliation{School of Physics, Huazhong University of Science and Technology, Wuhan 430074, China}
\affiliation{International Joint Laboratory on Quantum Sensing and Quantum Metrology, Huazhong University of Science and Technology, Wuhan 430074, China}
\author{Martin B. Plenio}
\affiliation{Institut f\"{u}r Theoretische Physik $\&$ IQST, Albert-Einstein Allee 11, Universit\"{a}t Ulm, D-89081, Germany}
\affiliation{International Joint Laboratory on Quantum Sensing and Quantum Metrology, Huazhong University of Science and Technology, Wuhan 430074, China}
\author{Fedor Jelezko}
\affiliation{Institut f\"{u}r Quantenoptik $\&$ IQST, Albert-Einstein Allee 11, Universit\"{a}t Ulm, D-89081, Germany}
\affiliation{International Joint Laboratory on Quantum Sensing and Quantum Metrology, Huazhong University of Science and Technology, Wuhan 430074, China}
\author{Tomoki Ozawa}
\affiliation{Interdisciplinary Theoretical and Mathematical Sciences Program (iTHEMS), RIKEN, Wako, Saitama 351-0198, Japan}
\author{Nathan Goldman}
\email{ngoldman@ulb.ac.be}
\affiliation{CENOLI, Universit\'e Libre de Bruxelles, CP 231, Campus Plaine, B-1050 Brussels, Belgium}
\author{Jianming~Cai}
\email{jianmingcai@hust.edu.cn}
\affiliation{School of Physics, Huazhong University of Science and Technology, Wuhan 430074, China}
\affiliation{International Joint Laboratory on Quantum Sensing and Quantum Metrology, Huazhong University of Science and Technology, Wuhan 430074, China}

\begin{abstract}
Geometry and topology are fundamental concepts, which underlie a wide range of fascinating physical phenomena such as topological states of matter and topological defects. In quantum mechanics, the geometry of quantum states is fully captured by the quantum geometric tensor. Using a qubit formed by an NV center in diamond, we perform the first experimental measurement of the complete quantum geometric tensor. Our approach builds on a strong connection between coherent Rabi oscillations upon parametric modulations and the quantum geometry of the underlying states. We then apply our method to a system of two interacting qubits, by exploiting the coupling between the NV center spin and a neighboring $^{13}$C nuclear spin. Our results establish coherent dynamical responses as a versatile probe for quantum geometry, and they pave the way for the detection of novel topological phenomena in solid state. 

\end{abstract}



\maketitle

{\it Introduction.---}  The quantum geometric tensor (QGT) constitutes a central and ubiquitous concept in quantum mechanics, by providing a geometric structure to the Hilbert space \cite{01,02,03,04,05}. The imaginary part of this tensor corresponds to the well-known Berry curvature \cite{06,07}, which acts as an effective ``electromagnetic'' tensor in parameter space. This geometric quantity, which is formally associated with the parallel transport of wave functions \cite{08}, is responsible for striking observable phenomena such as the geometric phase \cite{08}, the anomalous Hall effect \cite{09}, and topological states of matter \cite{10}. In contrast, the real part of the QGT constitutes the Fubini-Study metric \cite{02,03,05}, which defines a notion of distance (a Riemannian metric) in parameter space through the overlap of wavefunctions. This ``quantum metric'', which is intimately related to quantum fluctuations and dissipative responses of the system~\cite{02,05,11,12}, was shown to play an important role in various contexts, including quantum phase transitions \cite{13}, open quantum systems \cite{14}, orbital magnetism \cite{15,16}, localization in insulators \cite{11}, semiclassical dynamics \cite{17,18}, excitonic Lamb-shifts in transition-metal dichalcogenides \cite{19}, superfluidity in flat bands \cite{20}, and topological matter \cite{21,22}. In the context of quantum information, the quantum metric is equivalent to the quantum Fisher information, which is a witness for multipartite entanglement~\cite{23}.

Various manifestations of the QGT have been observed in experiments, using very different physical platforms and probes. On the one hand, the local Berry curvature has been detected in ultracold atomic gases \cite{24,25,26}, coupled optical fibers \cite{27}, and solids \cite{28,29}. On the other hand, a first manifestation of the quantum metric --- the so-called Wannier-spread functional of Bloch bands \cite{30} --- was recently measured in cold atoms \cite{31}, based on the proposal \cite{32}; see \cite{12, 33,34,35} for other proposals to detect quantum geometry. Nevertheless, a direct and systematic measurement of the complete QGT has never been performed.

Here, we report on the first experimental measurement of the complete QGT, using a qubit formed by an NV center spin in diamond. Following the proposal of Ref.~\cite{32}, we exploit the relation between the QGT and the response of quantum systems upon parametric modulations in order to map out the full Fubini-Study metric as well as the local Berry curvature of the underlying quantum states. We then apply our method to a system of two interacting qubits, obtained by coupling the NV center spin to a nearby $^{13}$C nuclear spin. Our results do not only enforce the deep connections between out-of-equilibrium dynamics and quantum geometry \cite{36,37,38,39,40,41,42,43,44}, but they also reveal a universal tool for the detection of geometric and topological properties in quantum systems.
%

%
\begin{figure}[t]
\begin{centering}
\includegraphics[width=8.8cm]{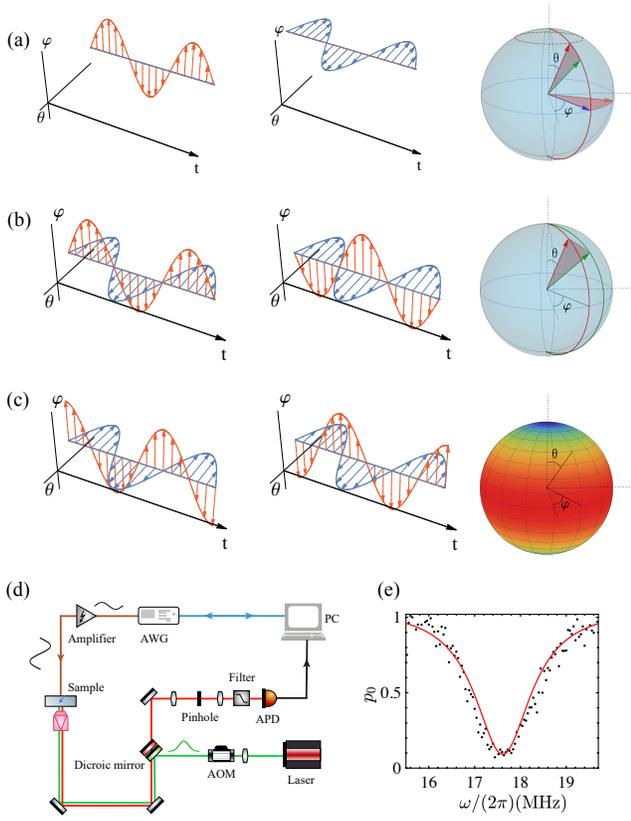}
\par\end{centering}
\caption{{\bf Probing quantum geometry through coherent responses upon parametric modulations}. (a-c) show different types of parametric modulations $(\theta_{t},\varphi_{t})$, including (a-b) linear parametric modulation $\theta_{t}=\theta_{0}+a_{\theta}\sin(\omega t),\varphi_{t}=\varphi_{0}+a_{\varphi}\sin(\omega t)$ for the measurement of the diagonal {[}off-diagonal{]} element of the Fubini-Study metric with $a_{\theta}=0$ or $a_{\varphi}=0$ (a) {[}$a_{\theta}=\pm a_{\varphi}\protect\neq0$ (b){]}; (c) elliptical parametric modulation for the measurement of the local Berry curvature (as indicated by color map) with $\theta_{t}=\theta_{0}+a_{\theta}\sin(\omega t)$ and $\varphi_{t}=\varphi_{0}\pm a_{\varphi}\cos(\omega t)$. (d) The sketch of experiment setup used for the quantum-geometric measurement, based on an NV center spin in diamond. A green laser pulse polarizes the NV center spin into the $|m_{s}=0\rangle$ state. The engineered microwave created from an arbitrary waveform generator (Tektronix AWG 70002A, 16 GS/s) is amplified before being delivered to the sample and coherently drives the NV center spin. The NV center spin state is detected by an APD via spin-dependent fluorescence. (e) An example of parametric-modulation resonance measurement. The probability that the NV center spin remains in the initial eigenstate at time $T=400$ ns as a function of the modulation frequency $\omega$ , for a linear parametric modulation $\theta_{t}=\theta_{0}+a_{\theta}\sin(\omega t),\varphi_{t}=\varphi_{0}$ with $(\theta_{0},\varphi_{0})=\left(\frac{5\pi}{6},0\right)$ and $a_{\theta}=0.1$. \label{fig:figure1}}
\end{figure}
%

{\it Detecting the QGT through Rabi oscillation.---} We start by considering the Hamiltonian $\mathcal{H}\left(\boldsymbol{\lambda}\right)$ of a generic discrete quantum system, which depends on a set of dimensionless parameters $\boldsymbol{\lambda}=(\lambda_{1},\lambda_{2},\cdots,\lambda_{N})$ , where $N$ is the dimension of parameter space. For a single qubit, the relevant parameter space corresponds to the two-dimensional Bloch sphere. Defining the eigenstates and eigenvalues of this generic Hamiltonian,
%
$\mathcal{H}\left(\boldsymbol{\lambda}\right)\left|n\left(\boldsymbol{\lambda}\right)\right\rangle =\epsilon_{n}\left(\boldsymbol{\lambda}\right)\left|n\left(\boldsymbol{\lambda}\right)\right\rangle$,
%
a geometric structure emerges upon projecting the dynamics onto a single (non-degenerate) band $\epsilon_{n}\left(\boldsymbol{\lambda}\right).$ The resulting quantum geometry is captured by the QGT, which is defined as \cite{07}
\begin{equation}
\chi_{\mu\nu}^{(n)}=\langle\partial_{\mu}n(\boldsymbol{\lambda})|\left(1-\right|n\left(\boldsymbol{\lambda}\right)\rangle\langle n(\boldsymbol{\lambda})|)|\partial_{\nu}n(\boldsymbol{\lambda})\rangle.
\end{equation}
For simplicity, hereafter we denote $ \partial_{\mu}\equiv  \partial_{\lambda_{\mu}}$. The real part $\text{Re}\left(\chi_{\mu\nu}\right)=g_{\mu\nu}$ is the Fubini-Study metric, which introduces a notion of distance in parameter space, while the imaginary part $\text{Im}\left(\chi_{\mu\nu}\right)=-\mathcal{F}_{\mu\nu}/2$ is related to the Berry curvature $\mathcal{F}_{\mu\nu}$ responsible for the Berry phase. It is useful to express the QGT in the form
\begin{equation}
\chi_{\mu\nu}^{(n)}=\sum_{m\neq n}\frac{\langle n\left(\boldsymbol{\lambda}\right)\left|\partial_{\mu}\mathcal{H}\left(\boldsymbol{\lambda}\right)\right|m\left(\boldsymbol{\lambda}\right)\rangle\langle m\left(\boldsymbol{\lambda}\right)\left|\partial_{\nu}\mathcal{H}\left(\boldsymbol{\lambda}\right)\right|n\left(\boldsymbol{\lambda}\right)\rangle}{\left(\epsilon_{m}\left(\boldsymbol{\lambda}\right)-\epsilon_{n}\left(\boldsymbol{\lambda}\right)\right)^{2}}\label{eq:chi}
\end{equation}
so as to highlight the relation between this geometric quantity and the coupling matrix elements connecting the eigenstates $|n(\boldsymbol{\lambda})\rangle$ and $|m(\boldsymbol{\lambda})\rangle$ upon a parametric modulation \cite{32},
\begin{equation}
\Omega_{n\leftrightarrow m}\left(\boldsymbol{\lambda}\right)\propto\langle m\left(\boldsymbol{\lambda}\right)\left|\partial_{\mu}\mathcal{H}\left(\boldsymbol{\lambda}\right)\right|n\left(\boldsymbol{\lambda}\right)\rangle.\label{eq:rabi}
\end{equation}
%


{\it Experimental setup.---}In our experiment, we first perform a full quantum-geometric measurement using a two-level system, as described by the general Hamiltonian
\begin{equation}
\mathcal{H}(\theta,\varphi)=\frac{A}{2}\begin{pmatrix}\cos\theta & \sin\theta e^{-i\varphi}\\
\sin\theta e^{i\varphi} & -\cos\theta
\end{pmatrix},\label{eq:hamiltonian}
\end{equation}
where the angles $(\theta,\varphi)$ form the relevant parameter space (the Bloch sphere). Considering the low-energy dressed state, the components of the QGT read
%
$g_{\ensuremath{\theta\theta}}=\frac{1}{4},\,\,g_{\ensuremath{\varphi\varphi}}=\frac{1}{4}\sin^{2}\theta,\,\,g_{\ensuremath{\theta\varphi}}=0,\,\,\mathcal{F}_{\ensuremath{\theta\varphi}}=\sin{\theta}/{2}$.
%
These components fully characterize the underlying quantum geometry: the quantum metric $g$ corresponds to the natural metric of a sphere $S^{2}$, embedded in $\mathbb{R}^{3}$ with fixed radius $R=\frac{1}{2}$, while the Berry curvature $\mathcal{F}_{\theta\varphi}$ corresponds to the ``magnetic'' field of a fictitious Dirac monopole located at the center of that sphere \cite{22}.

The experimental setup is sketched in Fig.\ref{fig:figure1}(d).  The two-level system in Eq.(\ref{eq:hamiltonian}) is obtained from a single nitrogen-vacancy (NV) center in an electronic grade diamond. We apply a magnetic field $B_{z}=509$ G along the NV axis to lift the degeneracy of the states $m_{s}=\pm1$. A two-level system is supported by the spin sublevels $m_{s}=0$ and $m_{s}=-1$. We first prepare the system in the eigenstate of the Hamiltonian $\mathcal{H}(\theta_{0},\varphi_{0})$, i.e. $|n\left(\theta_{0},\varphi_{0}\right)\rangle=\cos\frac{\theta_{0}}{2}|-1\rangle+\sin\frac{\theta_{0}}{2}e^{i\varphi_{0}}|0\rangle$. This is achieved by first applying a 532 nm green laser pulse to initialize the NV center spin in the $m_{s}=0$ state. A subsequent microwave pulse $\mathcal{H}_{I}(t)=\Omega\sin\left(\omega_{0}t+\varphi_{0}\right)\sigma_{x},$ applied over a duration $t_{\theta_{0}}=\frac{\theta_{0}}{\Omega}$, rotates the NV center spin around the axis $\hat{n}(\varphi_{0})=(\cos\varphi_{0},\sin\varphi_{0},0)$ by an angle $\theta_{0}$. The initial state preparation is verified by a spin-locking type experiment, which confirms that the NV spin is prepared in the eigenstate of $\mathcal{H}(\theta_{0},\varphi_{0})$ \cite{45}. 

The precise control over the AWG allows us to engineer the microwave driving field with accurate amplitude and phase modulation. This leads to the implementation of the generic two-level system
\begin{equation}
\mathcal{H}\left(t\right)=\frac{\omega_{0}}{2}\sigma_{z}+V\left(t\right)\sigma_{x},\label{eq:microwave}
\end{equation}
where $V\left(t\right)=(A\sin\theta_{t})\cos\left[\omega_{0}t-f\left(t\right)+\varphi_{t}\right]$. In the experiment, we calibrate the driving amplitude in the Hamiltonian Eq.\eqref{eq:microwave} with the output power of the AWG by measuring the Rabi frequency of the NV center spin \cite{45}. The amplitude modulation $A\sin\theta_{t}$ and the phase modulation $-f(t)+\varphi_{t}$ are synthesised by waveform programming in the AWG. The additional phase control function has the form $f\left(t\right)=A\int_{0}^{t}\cos\theta_{\tau}d\tau\cong A\cos\theta_{0}\mathcal{J}_{0}\left(a_{\theta}\right)t-\left(4A\sin\theta_{0}/\omega\right)\mathcal{J}_{1}\left(a_{\theta}\right)\sin^{2}(\omega t/2)$, where $\mathcal{J}_{0,1}$ are the zeroth and first order Bessel functions of the first kind, respectively \cite{45}. Taking the limit $\omega_{0}\gg A$, such an engineered microwave driving field allows us to realize the effective Hamiltonian in Eqs. (\ref{eq:hamiltonian}) 
with the designed parametric modulation \cite{45}:
\begin{equation}
\mathcal{H}_{\text{eff}}\left(t\right)\cong\frac{A}{2}\left\lbrack\cos\theta_{t}\sigma_{z}+\sin\theta_{t}\left(\cos\varphi_{t}\sigma_{x}+\sin\varphi_{t}\sigma_{y}\right)\right\rbrack .
\end{equation}
The parametric modulation drives a coherent transition between the eigenstates of $\mathcal{H}(\theta_{0},\varphi_{0})$, which is detected by rotating the NV center spin around the axis $\hat{n}(\varphi_{0})$ by an angle $2\pi-\theta_{0}$. This rotation maps the eigenstates of $\mathcal{H}(\theta_{0},\varphi_{0})$ back to the NV center spin state $|0\rangle$ and$|-1\rangle$, which is then measured by spin-dependent fluorescence.


%
\begin{figure}[t]
	\begin{centering}
		\includegraphics[width=8.8cm]{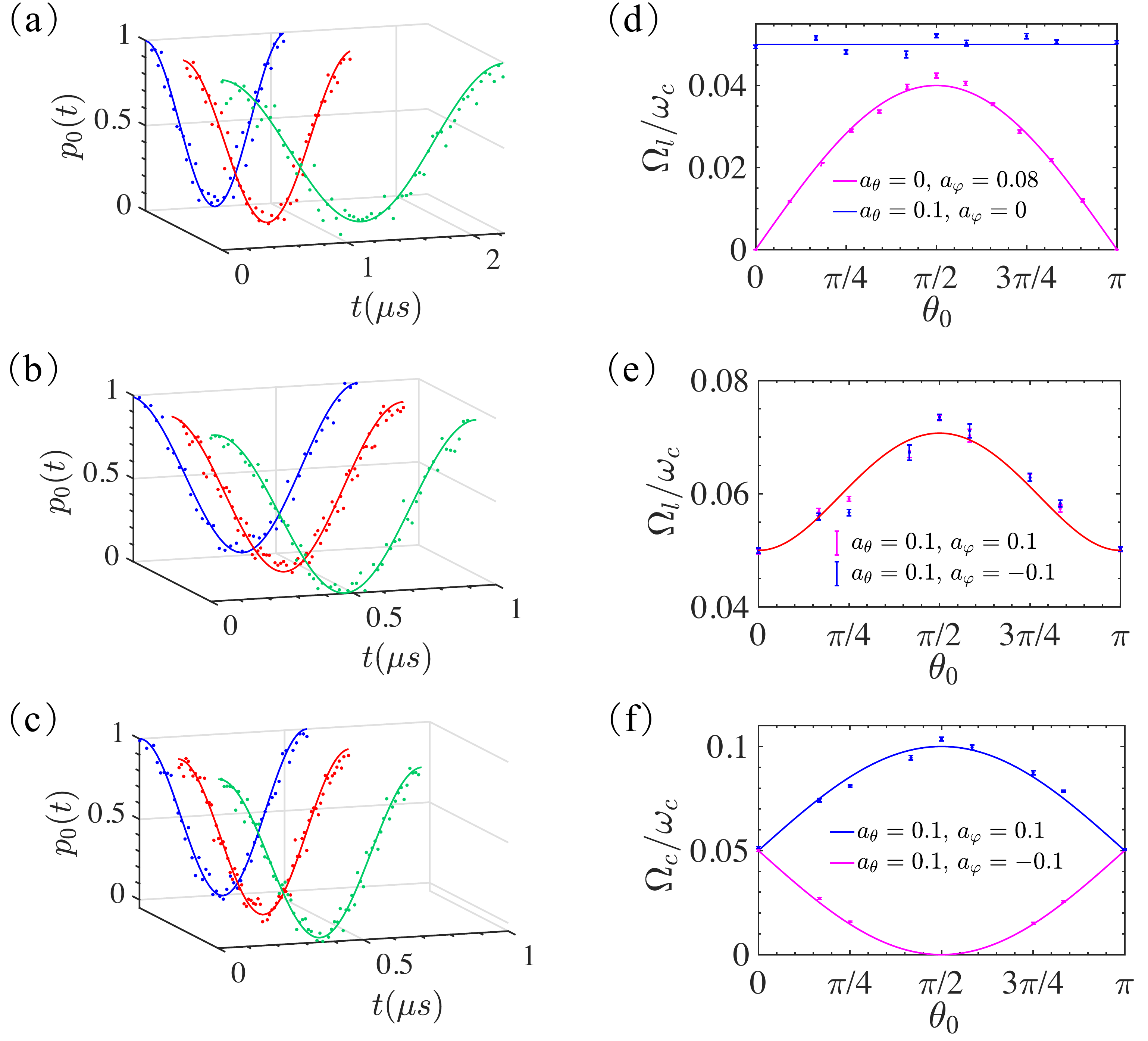}
		\par\end{centering}
	\caption{{\bf Coherent transitions induced by parametric modulations}. (a-b) Resonant oscillation under a linear parametric modulation with $a_{\theta}=0$, $a_{\varphi}=0.08$ (a) and $a_{\theta}=0.1$, $a_{\varphi}=0.1$ (b). (c) Resonant oscillation under an elliptical parametric modulation with $a_{\theta}=0.1$, $a_{\varphi}=0.1$. The other experimental parameters are: (a) $\omega_{c}=(2\pi)20.98$ MHz ($\theta_{0}=\frac{\pi}{6}$, green), $(2\pi)21.61$ MHz ($\theta_{0}=\frac{\pi}{3}$, red), $(2\pi)20.73$ MHz ($\theta_{0}=\frac{\pi}{2}$, blue); (b) $\omega_{c}=\left(2\pi\right)19.11$ MHz ($\theta_{0}=\frac{\pi}{6}$, green), $(2\pi)17.8$ MHz $(\theta_{0}=\frac{5\pi}{12}$ , red), $\left(2\pi\right)16.72$ MHz ($\theta_{0}=\frac{\pi}{2}$, blue); (c) $\omega_{c}=(2\pi)19.11$ MHz ($\theta_{0}=\frac{\pi}{6}$, green), $\left(2\pi\right)17.8$ MHz ($\theta_{0}=\frac{5\pi}{12}$, red), $\left(2\pi\right)16.72$ MHz ($\theta_{0}=\frac{\pi}{2}$, blue). (d-f) Rabi frequency of resonant coherent transitions upon parametric modulations (in the unit of resonant frequency $\omega_{c}$), as a function of the parameter $\theta_{0}$ , for linear (d-e) and elliptical (f) parametric modulations. The curves show theoretical predictions. In (a-f), we set the parameter $\varphi_{0}=$0. \label{fig:figure2}}
\end{figure}
%

{\it Experimental results.---} In the experiment, we implement two types of modulations \cite{32}: (a) a ``linear'' modulation $\theta_{t}=\theta_{0}+a_{\theta}\sin(\omega t)$, $\varphi_{t}=\varphi_{0}+a_{\varphi}\sin(\omega t)$; (b) an ``elliptical'' modulation $\theta_{t}=\theta_{0}+a_{\theta}\sin(\omega t)$, $\varphi_{t}=\varphi_{0}+a_{\varphi}\cos(\omega t)$; see Fig.\ref{fig:figure1}(a-c). Setting $a_{\theta},a_{\varphi}\ll1$, the time-dependent Hamiltonian can be expressed as
\begin{eqnarray}
\mathcal{H}(\theta_{t},\varphi_{t}) & \cong & \mathcal{H}(\theta_{0},\varphi_{0})+a_{\theta}(\partial_{\theta}\mathcal{H})\sin(\omega t)\nonumber \\
& + & a_{\varphi}(\partial_{\varphi}\mathcal{H})\sin(\omega t):\text{linear}\nonumber \\
& + & a_{\varphi}(\partial_{\varphi}\mathcal{H})\cos\left(\omega t\right):\text{elliptical}.\label{eq:time_dep}
\end{eqnarray}
%
After preparing the NV center spin in the eigenstate $|n\left(\theta_{0},\varphi_{0}\right)\rangle$ of the Hamiltonian $\mathcal{H}(\theta_{0},\varphi_{0})$, we apply the engineered microwave driving field with parametric modulation {[}see Eq.\eqref{eq:microwave}{]} and fix the time duration $T$. We sweep the parametric modulation frequency $\omega$, and measure the probability $p_{0}(T)$ that the NV spin remains in the initial eigenstate $|n\left(\theta_{0},\varphi_{0}\right)\rangle$. In Fig.\ref{fig:figure1} (e), we show an example of such a parametric-modulation resonance measurement; see Ref.\cite{45} for the experimental data using other types of modulations. The results indicate that a coherent transition between the eigenstates becomes resonant when $\omega \simeq A\equiv\omega_{c}$. We then measure the resonant coherent oscillation upon parametric modulation with $\omega$ =$\omega_{c}$, as shown in Fig.\ref{fig:figure2} (a-c). The observed Rabi frequencies under resonant parametric modulations, which reveal the information about the coupling matrix elements connecting the eigenstates (see Eq.\ref{eq:rabi}) upon a parametric modulation, are shown in Fig.\ref{fig:figure2}(d-f). The experiment results allow us to determine the quantum geometry of the prepared dressed states precisely.

%
\begin{figure}[b]
\begin{centering}
\includegraphics[width=1\columnwidth]{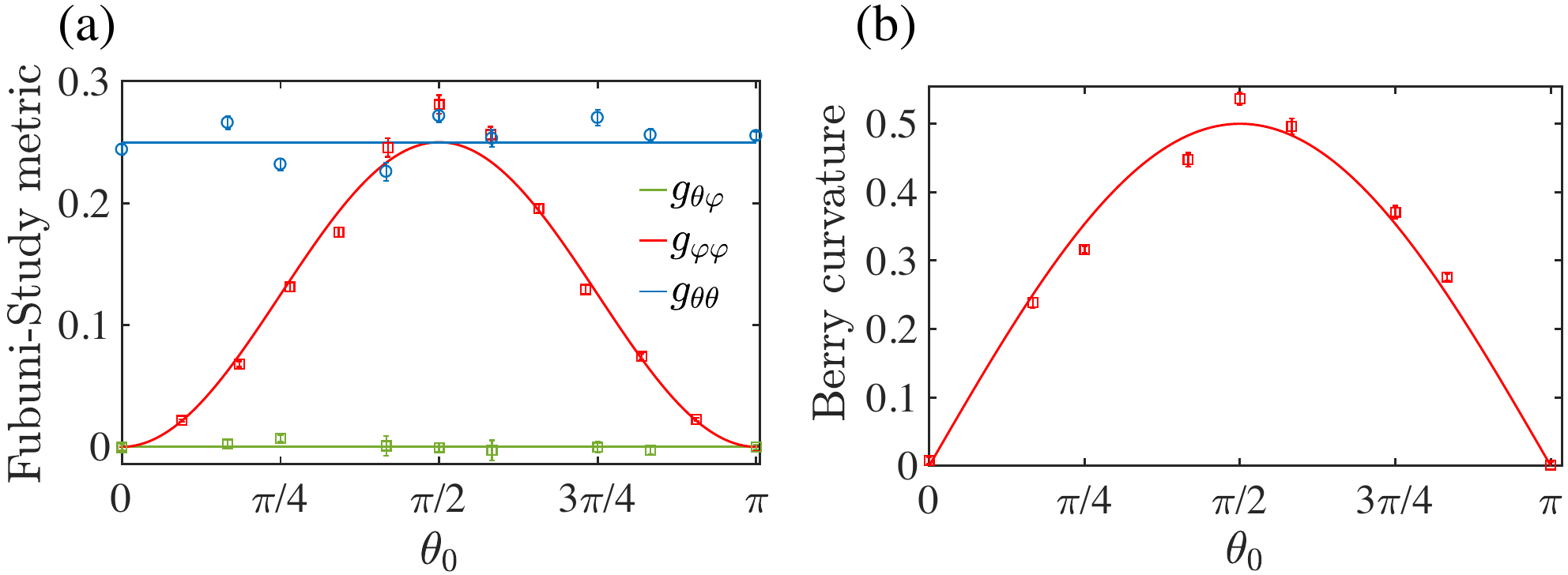}
\par\end{centering}
\caption{{\bf Extraction of the complete quantum geometric tensor}. (a) The measured Fubini-Study metric, compared with the theoretical predictions $g_{\theta\varphi}=0$ (green curve), $g_{\varphi\varphi}=\sin^{2}\theta_{0}/4$ (red curve) and $g_{\theta\theta}=1/4$ (blue curve). (b) The measured local Berry curvature $\mathcal{F}_{\theta\varphi}$ is compared with the theoretical prediction $\mathcal{F}_{\theta\varphi}=\sin{\theta_{0}}/{2}$. The experimental parameters are the same as in Fig.\ref{fig:figure2}. \label{fig:figure3}}
\end{figure}
%
%

As a central result, we show in Fig.\ref{fig:figure3} the experimental extraction of the full QGT, based on Rabi-oscillation measurements. This provides a first demonstration that coherent responses upon parametric modulations can be used as a powerful tool to access the complete geometry of a discrete quantum system. We point out that the present quantum-geometry measurement is based on coherent dynamical responses upon periodic driving, and in this sense, it does not rely on any adiabaticity constraints (i.e. small modulation velocity \cite{36,37}). It should be noted, however, that this method uses small modulation amplitudes, and hence small Rabi frequencies, which requires systems exhibiting long coherence times. The agreement between the experiment results and the theoretical predictions can be improved by increasing the measurement time, which allows for a better determination of the oscillation frequency. Furthermore, in contrast with the excitation-rate measurement of Refs. \cite{31,32,41}, the QGT is extracted from Rabi oscillations \cite{39}, where the initial state is recovered after each Rabi period; in principle, this allows for the detection of geometry and topology through a non-destructive measurement.

\begin{figure}[t]
\begin{centering}
\includegraphics[height=3cm]{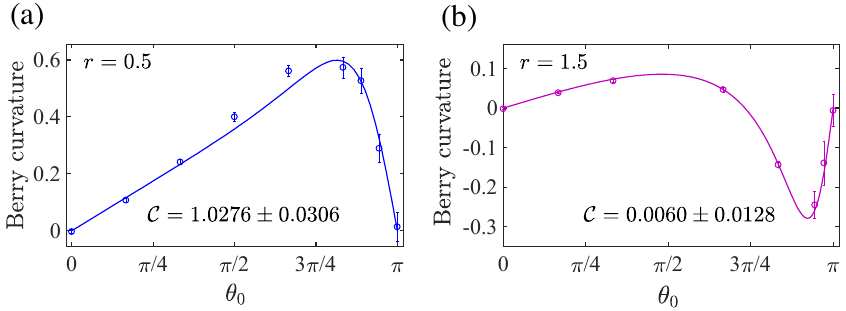}
\par\end{centering}
\caption{{\bf Berry curvature measurement across the topological transition}. (a-b) show the measured local Berry curvature $F_{\theta\varphi}$ for the Hamiltonian in Eq.\eqref{eq:hamiltonian_r}, which describes a Dirac monopole located inside ($a$, $r=0.5$) and outside ($b$, $r=1.5$) the Bloch sphere. The curves represent the corresponding theoretical values. The Chern number extracted from the data is indicated in both panels. \label{fig:figure4}}
\end{figure}

Besides, our quantum-geometry measurement can also be used to characterize the topology of the underlying system. For this analysis, we extend the Hamiltonian to the form

\begin{equation}
\mathcal{H}\left(\theta,\varphi\right)=\frac{A}{2}\begin{pmatrix}\cos\theta+r & \sin\theta e^{-i\varphi}\\
\sin\theta e^{i\varphi} & -\cos\theta-r
\end{pmatrix}\label{eq:hamiltonian_r}
\end{equation}
where $r$ is a tunable parameter. As for Eq.\eqref{eq:hamiltonian}, the geometry of the Hamiltonian in Eq.\eqref{eq:hamiltonian_r} is that of a fictitious monopole located close to a sphere $S^{2}$, whose position in parametric space depends on the additional parameter $r$. The topology of the system then relies on whether this fictitious monopole is located inside the sphere or not, as captured by the Chern number $C=\frac{1}{2\pi}\int_{S^{2}}^{} \mathcal{F}_{\theta\varphi}d\theta d\varphi$ \cite{13}. Fig.\ref{fig:figure4} shows the Berry-curvature measurement in two distinct topological phases. In the non-trivial regime, the Chern number can equally be determined from the metric $C=\frac{1}{2\pi}\int_{S^{2}}^ {}\left(2\sqrt{\overline{g}}\right)d\theta d\varphi=\frac{1}{2\pi}\int_{S^{2}}^ {}|\mathcal{F}_{\theta\varphi}|d\theta d\varphi$, where $\overline{g}=g_{\theta\theta}g_{\varphi\varphi}-(g_{\theta\varphi})^{2}$ is the determinant of the QGT \cite{22}. Altogether, these results indicate that topology can indeed be finely analyzed based on our geometric-detection scheme.

%
%
%

\begin{figure}[t]
	\begin{centering}
		\includegraphics[height=3cm]{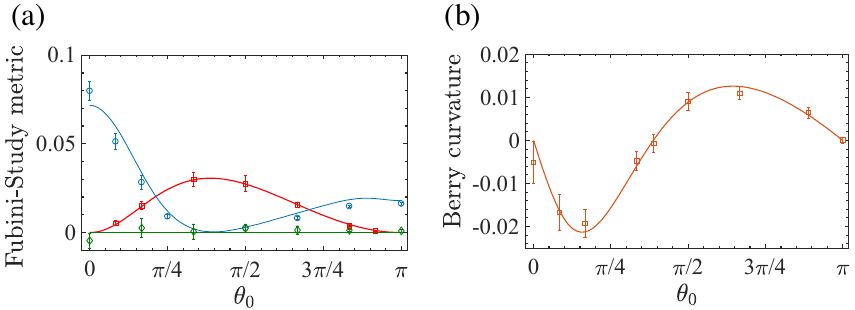}
	\end{centering}
	\caption{{\bf Quantum geometry of an interacting two-qubit system}. (a) The measured Fubini-Study metric, compared with the theoretical predictions: $g_{\theta\varphi}$ (green curve), $g_{\varphi\varphi}$ (red curve) and $g_{\theta\theta}$(blue curve). (b) The measured local Berry curvature $\mathcal{F}_{\theta\varphi}$ is compared with the theoretical prediction (curve). The amplitude of the driving field [see Eq.\eqref{eq:Hrot-two-qubit}] is $\Omega_{\text{mw}}=2.13$ MHz. The Chern number estimated from the integral of the Berry curvature is $C=0.0009\pm0.0067$, which is in agreement with the prediction ($C=0$) in this strongly-interacting regime. \label{fig:figure5}}
\end{figure}

{\it Application to interacting qubits.---} As a second application, we further extend our experiment to extract the QGT of an interacting two-qubit system. The interacting two-qubit system is formed by an NV center electron spin coupled to a $^{13}$C nuclear spin located in the vicinity of the NV center. We have determined the strength of the corresponding spin-spin interactions using a pulsed optically detected magnetic resonance (ODMR) experiment; we obtain the interaction parameters: $A_{x}\approx 2.79$ MHz and $A_{z}\approx11.832$ MHz [see Eq.\eqref{eq:Hrot-two-qubit} below]. By engineering microwave driving fields with designed frequency and phase, we obtain the following effective Hamiltonian
\begin{eqnarray}
\nonumber 
\mathcal{H}_{\text{rot}}\left(\theta,\varphi\right)=&&\frac{\Omega_{\text{mw}}}{2}\left\lbrack \cos\theta\sigma_{z}+\sin\theta\left(\cos\varphi\sigma_{x}+\sin\varphi\sigma_{y}\right)\right\rbrack\\ \nonumber 
 &&+\left(\frac{\gamma_{n}B_{\parallel}}{2}-\frac{A_{z}}{4}\right)\tau_{z}-\frac{A_{x}}{4}\tau_{x}\\
&&-\frac{A_{z}}{4}\sigma_{z}\otimes\tau_{z}-\frac{A_{x}}{4}\sigma_{z}\otimes\tau_{x} , \label{eq:Hrot-two-qubit}
\end{eqnarray}
where $\boldsymbol{\sigma}$ and $\boldsymbol{\tau}$ are Pauli operators associated with the first and second qubit, respectively. Henceforth, we denote the eigenstates of the Hamiltonian in Eq.\eqref{eq:Hrot-two-qubit} as $\ket{\Psi_1}$, $\ket{\Psi_2}$, $\ket{\Psi_3}$, $\ket{\Psi_4}$, according to their ordered eigenenergies $\epsilon_1< \epsilon_2<\epsilon_3 <\epsilon_4$. 
The competition between the local term ($\Omega_{\text{mw}}$) and the spin-spin interaction in the Hamiltonian Eq.\eqref{eq:Hrot-two-qubit} leads to a rich topological phase diagram. In the regime $\Omega_{\mathrm{mw}}\gg\Omega_{\mathrm{mw}}^{(c_1)}$, where 
\begin{equation}
\Omega_{\mathrm{mw}}^{(c_1)}=\frac{1}{2}\blb{-\gamma_nB_{\parallel}+\sqrt{(\gamma_nB_{\parallel}-A_{z})^2+A^2_{x}}} ,
\end{equation}
the spin-spin interaction becomes less significant and we thus recover the topological properties of the two-level system, for which the Chern number is $C\!=\!1$ in the eigenstate $\ket{\Psi_3}$ (see the measurements described in the previous section); note that the other eigenstates exhibit similar behaviours. The spin-spin interaction eventually dominates upon decreasing the value of the local parameter; below the critical value, $\Omega_{\mathrm{mw}}<\Omega_{\mathrm{mw}}^{(c_1)}$, the Chern number of the eigenstate changes from $C\!=\!1$ to $C\!=\!0$, which can be seen as a drastic effect of the spin-spin interaction. This vanishing of the Chern number in the strongly-interacting regime is clearly captured by our QGT measurement, as reported in Fig.\ref{fig:figure5}. These results demonstrate the measurement of both the Fubini-Study metric and the Berry curvature deep in the interacting regime, and show excellent agreement with theoretical predictions \cite{45}. 

%
%
%

As previously noted, the QGT contains information regarding the entanglement properties of interacting systems, through the concept of quantum Fisher information~\cite{23}. As an interesting perspective, our detection method could be applied to more complex interacting systems in view of revealing their quantum fluctuations and entanglement properties.

{\it Conclusion.---} To summarize, we have experimentally demonstrated a powerful connection between the quantum geometric tensor and the coherent dynamical response of a quantum system upon a parametric drive. Based on this fundamental relation, we have first extracted the complete QGT, including all the components of the Fubini-Study metric and those of the local Berry curvature, by driving Rabi oscillations in a single qubit. These measurements have clearly revealed the topological (monopole-type) structure associated with this simple setting. We point out that this method is readily applicable to observe other intriguing topological defects, such as tensor monopoles defined in 4D parameter spaces~\cite{22}. Furthermore, we have applied this detection method to an interacting two-qubit system, which suggests potential applications to many-body quantum systems with geometric features \cite{12, 32, 46}. Altogether, our results demonstrate that coherent dynamical responses can serve as a powerful tool to access the geometric and topological properties of quantum systems and open a way to explore the fundamental role of the QGT in various scenarios, ranging from many-body systems to open quantum systems.

{\it Acknowledgements.---} We thank G. Palumbo, M. Di Liberto, P. Zoller, Yu Liu for helpful
discussions. The work is supported by National Natural Science Foundation of China (11574103,
11690030, 11690032, 11874024). M.B.P. is supported by the EU projects HYPERDIAMOND and
AsteriQs, the BMBF project NanoSpin and DiaPol and the ERC Synergy grant BioQ. F. J. is
supported by DFG (FOR 1493, SPP 1923), VW Stiftung, BMBF, ERC, EU (AsteriQs), BW
Stiftung, Ministry of Science and Arts, Center for Integrated quantum science and technology
(IQST). T.O. is supported by JSPS KAKENHI Grant Number JP18H05857, RIKEN Incentive
Research Project, and the Interdisciplinary Theoretical and Mathematical Sciences Program
(iTHEMS) at RIKEN. N.G. is supported by the ERC Starting Grant TopoCold and the Fonds De
La Recherche Scientifique (FRS-FNRS) (Belgium).

{\it Note added.---} M. Y. and P. Y. contributed equally to this work. Two other experimental measurements of the QGT were reported after the completion of our work~\cite{47}, in polaritons \cite{48} and superconducting qubits \cite{49}.

 

\section*{Supplementary material}

# Supplementary Information for:
# "Experimental measurement of the quantum geometric tensor using coupled qubits in diamond"

## A. CONNECTION BETWEEN QUANTUM GEOMETRY AND RABI OSCILLATION

### A.1 Parametric modulation induced coherent transition for discrete quantum systems

In the main text, we experimentally demonstrate the connection between parametric modulation induced coherent transition and quantum geometric tensor for discrete quantum systems. Here, we provide a detailed analysis on such a fundamental connection following Ref.[1]. Consider a discrete quantum system with a general Hamiltonian $\mathcal{H}(\boldsymbol{\lambda})$ which depends on a set of dimensionless parameters $\boldsymbol{\lambda} = (\lambda_1, \lambda_2, \cdots, \lambda_N)$ where $N$ is the dimension of parameter space, the eigenstates are given as follows

$$\mathcal{H}(\boldsymbol{\lambda})|n(\boldsymbol{\lambda})\rangle = \epsilon_n(\boldsymbol{\lambda})|n(\boldsymbol{\lambda})\rangle. \tag{1}$$

If the Hamiltonian has no energy degeneracy, the definition of quantum geometric tensor (QGT) is [2]

$$\chi_{\mu\nu} = \langle \partial_\mu n(\boldsymbol{\lambda}) | (1 - |n(\boldsymbol{\lambda})\rangle\langle n(\boldsymbol{\lambda})|) | \partial_\nu n(\boldsymbol{\lambda}) \rangle. \tag{2}$$

The real part of QGT $\mathrm{Re}(\chi_{\mu\nu}) = g_{\mu\nu}$ is the Fubini-Study metric that quantifies the distance between nearby states $|n(\boldsymbol{\lambda})\rangle$ and $|n(\boldsymbol{\lambda} + d\boldsymbol{\lambda})\rangle$ on parametric manifold, the imaginary part $\mathrm{Im}(\chi_{\mu\nu}) = -\mathcal{F}_{\mu\nu}(\boldsymbol{\lambda})/2$ where $\mathcal{F}_{\mu\nu}(\boldsymbol{\lambda})$ is the local Berry curvature, which is responsible for the geometric (Berry) phase. The QGT is shown to connect with the coherent response on parametric modulation [1]. In the main text, we consider two types of parametric modulation $[\lambda_\mu(t), \lambda_\nu(t)]$: (a) the linear parametric modulation with $\lambda_\mu(t) = \lambda_\mu^0 + a_\mu \sin(\omega t)$, $\lambda_\nu(t) = \lambda_\nu^0 + a_\nu \sin(\omega t)$; (b) the elliptical parametric modulation with $\lambda_\mu(t) = \lambda_\mu^0 + a_\mu \sin(\omega t)$, $\lambda_\nu(t) = \lambda_\nu^0 + a_\nu \cos(\omega t)$.

As a specific example of linear parametric modulation with $a_\mu \neq 0$ and $a_\nu = 0$, when considering weak parametric modulation i.e. $a_\mu \ll 1$, the time-dependent Hamiltonian can be expanded as

$$\mathcal{H}[\boldsymbol{\lambda}(t)] = \mathcal{H}(\boldsymbol{\lambda}^0) + a_\mu \left[\partial_\mu \mathcal{H}(\boldsymbol{\lambda}^0)\right] \sin(\omega t). \tag{3}$$

Assume that the initial state is prepared on the ground state $|n(\boldsymbol{\lambda}^0)\rangle$, it will be excited onto the higher energy levels by the time-dependent Hamiltonian. Since the Hamiltonian is periodic, the problem can be solved by Floquet theorem. But if the frequency of modulation is on resonance with the energy detuning between $|n(\boldsymbol{\lambda}^0)\rangle$ and $|m(\boldsymbol{\lambda}^0)\rangle$ namely $\hbar\omega = \hbar\omega_{n\leftrightarrow m} \equiv |\epsilon_m(\boldsymbol{\lambda}) - \epsilon_n(\boldsymbol{\lambda})|$, the Floquet Hamiltonian can be simplified as a two-level Hamiltonian in the Hilbert subspace spanned by $\{|n(\boldsymbol{\lambda}^0)\rangle, |m(\boldsymbol{\lambda}^0)\rangle\}$ as follows

$$\mathcal{H}_{\mathrm{rot}}(\boldsymbol{\lambda}^0) = \begin{pmatrix} \epsilon_m(\boldsymbol{\lambda}) - \hbar\omega & \Omega_{n\leftrightarrow m}(\boldsymbol{\lambda}) \\ \Omega^*_{n\leftrightarrow m}(\boldsymbol{\lambda}) & \epsilon_n(\boldsymbol{\lambda}) \end{pmatrix}, \tag{4}$$

where

$$\Omega_{n\leftrightarrow m}(\boldsymbol{\lambda}) = \frac{a_\mu}{2} \langle m(\boldsymbol{\lambda}) | \partial_\mu \mathcal{H}(\boldsymbol{\lambda}) | n(\boldsymbol{\lambda}) \rangle, \tag{5}$$

with the following relation as

$$\begin{aligned} &\left|\langle m(\boldsymbol{\lambda})|\partial_\mu \mathcal{H}(\boldsymbol{\lambda})|n(\boldsymbol{\lambda})\rangle\right|^2 \\ &= [\epsilon_m(\boldsymbol{\lambda}) - \epsilon_n(\boldsymbol{\lambda})]^2 \langle \partial_\mu n(\boldsymbol{\lambda})|m(\boldsymbol{\lambda})\rangle\langle m(\boldsymbol{\lambda})|\partial_\mu n(\boldsymbol{\lambda})\rangle. \end{aligned} \tag{6}$$

According to the definition of QGT, it can be seen that the diagonal element of the Fubini-Study metric can be expressed as

follows

$$g_{\mu\mu}(\lambda) = \langle \partial_\mu n(\lambda)|(1 - |n(\lambda)\rangle\langle n(\lambda)|)|\partial_\mu n(\lambda)\rangle \tag{7}$$

$$= \sum_{m \neq n} \langle \partial_\mu n(\lambda)|m(\lambda)\rangle\langle m(\lambda)|\partial_\mu n(\lambda)\rangle \tag{8}$$

$$= \sum_{m \neq n} \frac{|\langle m(\lambda)|\partial_\mu \mathcal{H}(\lambda)|n(\lambda)\rangle|^2}{(\epsilon_m(\lambda) - \epsilon_n(\lambda))^2} \tag{9}$$

$$= \sum_{m \neq n} \frac{4|\Omega_{n \leftrightarrow m}(\lambda)|^2}{a_\mu^2 \omega_{n \leftrightarrow m}^2}. \tag{10}$$

For a two-level quantum system, the Rabi frequency of coherence transition is

$$\Omega_l(a_\mu) = 2|\Omega_{n \leftrightarrow m}(\lambda)| = a_\mu g_{\mu\mu}^{1/2}(\lambda)\omega_c, \tag{11}$$

where $\omega_c = \omega_{g \leftrightarrow e}$ represents the resonant frequency between the ground state and the excited state.

We proceed to consider general linear parametric modulation $\lambda_\mu(t) = \lambda_\mu^0 + a_\mu \sin(\omega t)$, $\lambda_\nu(t) = \lambda_\nu^0 + a_\nu \sin(\omega t)$ with $a_\mu \neq 0$ and $a_\nu \neq 0$. The amplitude of parametric modulation is small $|a_\mu|, |a_\nu| \ll 1$, thus the time-dependent Hamiltonian can be written as follows

$$\mathcal{H}[\lambda(t)] = \mathcal{H}(\lambda^0) + a_\mu \left[\partial_\mu \mathcal{H}(\lambda^0)\right] \sin(\omega t) \tag{12}$$

$$+ a_\nu \left[\partial_\nu \mathcal{H}(\lambda^0)\right] \sin(\omega t). \tag{13}$$

And the coherent transition Rabi frequency is

$$\Omega_{n \leftrightarrow m}(\lambda) = \frac{1}{2}\langle m(\lambda)|a_\mu \partial_\mu \mathcal{H}(\lambda) + a_\nu \partial_\nu \mathcal{H}(\lambda)|n(\lambda)\rangle. \tag{14}$$

Similarly, we can get the following relation between parametric modulation induced coherent transition and the Fubini-Study metric as

$$\sum_{m \neq n} \frac{4|\Omega_{n \leftrightarrow m}(\lambda)|^2}{\omega_{n \leftrightarrow m}^2} = a_\mu^2 g_{\mu\mu} + 2a_\mu a_\nu g_{\mu\nu} + a_\nu^2 g_{\nu\nu}. \tag{15}$$

In particular, for a two-level quantum system, the corresponding Rabi frequency of coherent transition $\Omega_l(a_\mu, a_\nu) = 2|\Omega_{n \leftrightarrow m}(\lambda)|$ is related to the Fubini-Study metric as follows

$$\Omega_l(a_\mu, a_\nu)^2 / \omega_{g \leftrightarrow e}^2 = a_\mu^2 g_{\mu\mu} + 2a_\mu a_\nu g_{\mu\nu} + a_\nu^2 g_{\nu\nu}. \tag{16}$$

Therefore, one can extract the off-diagonal element of the Fubini-Study metric as follows

$$g_{\mu\nu} = \left[\Omega_l(a_\mu, a_\nu)^2 - \Omega_l(a_\mu, -a_\nu)^2\right] / \left(4 a_\mu a_\nu \omega_{g \leftrightarrow e}^2\right). \tag{17}$$

The coherent response on the elliptical parametric modulation with $\lambda_\mu(t) = \lambda_\mu^0 + a_\mu \sin(\omega t)$, $\lambda_\nu(t) = \lambda_\nu^0 + a_\nu \cos(\omega t)$ can be analysed in a similar way. The corresponding time-dependent Hamiltonian can be written as follows

$$\mathcal{H}[\lambda(t)] = \mathcal{H}(\lambda^0) + a_\mu \left[\partial_\mu \mathcal{H}(\lambda^0)\right] \sin(\omega t)$$

$$+ a_\nu \left[\partial_\nu \mathcal{H}(\lambda^0)\right] \cos(\omega t). \tag{18}$$

The coherent transition Rabi frequency is

$$\Omega_{n \leftrightarrow m}(\lambda) = \frac{1}{2}\langle m(\lambda)|a_\mu \partial_\mu \mathcal{H}(\lambda) - i a_\nu \partial_\nu \mathcal{H}(\lambda)|n(\lambda)\rangle, \tag{19}$$

which is connected with the local Berry curvature as

$$\sum_{m \neq n} \frac{4|\Omega_{n \leftrightarrow m}(\lambda)|^2}{\omega_{n \leftrightarrow m}^2} = a_\mu^2 g_{\mu\mu} + a_\mu a_\nu \mathcal{F}_{\mu\nu} + a_\nu^2 g_{\nu\nu}. \tag{20}$$

For a two-level quantum system, we have the Rabi frequency of coherent transition $\Omega_c(a_\mu, a_\nu) = 2|\Omega_{n\leftrightarrow m}(\lambda)|$ is related to the Fubini-Study metric and the local Berry curvature as follows

$$\Omega_c(a_\mu, a_\nu)^2/\omega_{g\leftrightarrow e}^2 = a_\mu^2 g_{\mu\mu} + a_\mu a_\nu \mathcal{F}_{\mu\nu} + a_\nu^2 g_{\nu\nu}. \tag{21}$$

Therefore, one can measure the local Berry curvature in the following way

$$\mathcal{F}_{\mu\nu} = \left[\Omega_c(a_\mu, a_\nu)^2 - \Omega_c(a_\mu, -a_\nu)^2\right] / \left(2 a_\mu a_\nu \omega_{g\leftrightarrow e}^2\right). \tag{22}$$

We remark that, for a discrete quantum system with more than two energy levels, the frequency shall be modulated in a larger range and a series of resonant transitions from the specific eigenstate to the other eigenstates shall be taken into account respectively. In the limit $a_\mu, a_\nu \ll 1$ where $\mu, \nu \in \{\lambda_1, \lambda_2, \cdots, \lambda_N\}$, each resonant transition can be approximated as a two-level system and the transition element $\langle m(\lambda)|\partial_\mu \mathcal{H}(\lambda)|n(\lambda)\rangle$ [see Eq.(3) in the main text] can be measured similarly.

### A.2 Floquet analysis of coherent response on parametric modulation

The Hamiltonian of a two-level quantum system that we realize in the main text is given in Eq.(5). The system's response to periodic parametric modulation reveals information on quantum geometry. Here, we provide Floquet analysis for the example of linear parametric modulation $\theta(t) = \theta_0 + a_\theta \sin(\omega t)$, $\varphi(t) = \varphi_0 + a_\varphi \sin(\omega t)$ with $a_\theta = 0$ and $a_\varphi \neq 0$. In this case, the time-dependent Hamiltonian can be expanded as

$$\mathcal{H}(t) = \mathcal{H}_0(\theta_0, \varphi_0) + \sum_{n\neq 0} \mathcal{H}_n e^{in\omega t}, \tag{23}$$

with

$$\mathcal{H}_n = \frac{A}{2} \mathcal{J}_n(a_\varphi) \sin\theta_0 \begin{pmatrix} 0 & e^{-i\varphi_0} \\ (-1)^n e^{i\varphi_0} & 0 \end{pmatrix}, \tag{24}$$

where $\mathcal{J}_n$ is the n-th order Bessel function of the first kind. The eigenstates of the Hamiltonian $\mathcal{H}_0(\theta_0, \varphi_0)$ are

$$|\psi_1\rangle = \cos\frac{\theta_0}{2}|\uparrow\rangle + \sin\frac{\theta_0}{2} e^{i\varphi_0}|\downarrow\rangle, \tag{25}$$

$$|\psi_2\rangle = -\sin\frac{\theta_0}{2} e^{-i\varphi_0}|\uparrow\rangle + \cos\frac{\theta_0}{2}|\downarrow\rangle, \tag{26}$$

with the corresponding eigenenergy $\pm\frac{A}{2}$ respectively. Then one can rotate the above time-dependent Hamiltonian written in the basis $\{|\psi_1\rangle, |\psi_2\rangle\}$ as

$$\mathcal{H}'(t) = \frac{A}{2}(|\psi_1\rangle\langle\psi_1| - |\psi_2\rangle\langle\psi_2|) + \sum_{n\neq 0} \mathcal{H}'_n e^{in\omega t}, \tag{27}$$

where

$$\mathcal{H}'_{2n-1} = \frac{A}{2} \mathcal{J}_{2n-1}(a_\varphi) \sin\theta_0 \begin{pmatrix} 0 & e^{-i\varphi_0} \\ -e^{i\varphi_0} & 0 \end{pmatrix}, \tag{28}$$

$$\mathcal{H}'_{2n} = \frac{A}{2} \mathcal{J}_{2n}(a_\varphi) \sin\theta_0 \begin{pmatrix} \sin\theta_0 & \cos\theta_0 e^{-i\varphi_0} \\ \cos\theta_0 e^{i\varphi_0} & -\sin\theta_0 \end{pmatrix}. \tag{29}$$

Under rotating wave approximation, in the resonance case $\hbar\omega \cong A$, the higher order terms can be ignored and the above Hamiltonian can be simplified as

$$\mathcal{H}'(t) = \frac{A}{2}(|\psi_1\rangle\langle\psi_1| - |\psi_2\rangle\langle\psi_2|) \tag{30}$$
$$+ \frac{A}{2} \mathcal{J}_1(a_\varphi) \sin\theta_0 e^{i\omega t} \left(e^{-i\varphi_0}|\psi_1\rangle\langle\psi_2| - e^{i\varphi_0}|\psi_2\rangle\langle\psi_1|\right) + h.c.$$

which leads to the following effective Hamiltonian as

$$\mathcal{H}_{\rm rot} = \begin{pmatrix} \frac{A}{2} & -\frac{A}{2}\mathcal{J}_1(a_\varphi)\sin\theta_0 e^{i\varphi_0} \\ -\frac{A}{2}\mathcal{J}_1(a_\varphi)\sin\theta_0 e^{-i\varphi_0} & -\frac{A}{2} + \hbar\omega \end{pmatrix}. \tag{31}$$

Therefore, the Rabi frequency of coherent transition from the ground state to the excited state is

$$\Omega = \frac{A}{2}\mathcal{J}_1(a_\varphi)\sin\theta_0 = a_\varphi g_{\varphi\varphi}^{1/2}\omega_c/2. \tag{32}$$

We remark that the above Floquet analysis can be extended to general linear parametric modulation as well as elliptical parametric modulation.

### A.3 Engineering of the effective Hamiltonian with parametric modulation

In the main text, we engineer a microwave driving field with amplitude, frequency and phase modulation [see Eq.(5)] acting on the NV center spin that leads to the following Hamiltonian as

$$\mathcal{H}(t) = \frac{\omega_0}{2}\sigma_z + A\sin\theta_t \cos\left[\omega_0 t - f(t) + \varphi_t\right]\sigma_x, \tag{33}$$

where $\theta_t$ and $\varphi_t$ are periodically modulated parameters and $f(t)$ is a function which can be controlled in experiment. We first define the operator

$$\mathcal{K}(t) = \frac{\omega_0 t}{2}\sigma_z - \frac{A}{2}\int_0^t \cos\theta_\tau d\tau\,\sigma_z \equiv \mathcal{A}(t)\sigma_z. \tag{34}$$

The effective Hamiltonian in the rotating frame can be written as

$$\begin{aligned}
\mathcal{H}_{\text{eff}} &= e^{i\mathcal{K}(t)}\mathcal{H}(t)e^{-i\mathcal{K}(t)} + i\left(\frac{\partial e^{i\mathcal{K}(t)}}{\partial t}\right)e^{-i\mathcal{K}(t)} \\
&\cong \frac{A}{2}\sin\theta_t \cos\left[A\int_0^t \cos\theta_\tau d\tau - f(t) + \varphi_t\right]\sigma_x \\
&\quad + A\sin\theta_t \sin\left[A\int_0^t \cos\theta_\tau d\tau - f(t) + \varphi_t\right]\sigma_y \\
&\quad + \frac{A}{2}\cos\theta_t \sigma_z,
\end{aligned} \tag{35}$$

where we use rotating-wave approximation and neglect the fast oscillating terms with frequency $2\omega_0$. If we set the phase control function $f(t) = A\int_0^t \cos\theta_\tau d\tau$, the effective Hamiltonian can be simplified as

$$\mathcal{H}_{\text{eff}} \cong \frac{A}{2}\left[\cos\theta_t\sigma_z + \sin\theta_t\left(\cos\varphi_t\sigma_x + \sin\varphi_t\sigma_y\right)\right] \tag{36}$$

which is the desired Hamiltonian. In our experiments, we implement the phase control function $f(t)$ for $\theta_t = \theta_0 + a_\theta \sin(\omega t)$ as follows

$$\begin{aligned}
f(t) &= A\int_0^t \cos\theta_\tau d\tau \\
&= A\int_0^t \cos\left[\theta_0 + a_\theta \sin(\omega\tau)\right] d\tau \\
&= A\int_0^t \left\{\cos\theta_0\left[\mathcal{J}_0(a_\theta) + 2\sum_{n=1}^\infty \mathcal{J}_{2n}(a_\theta)\cos(2n\omega\tau)\right]\right. \\
&\quad \left. - 2\sin\theta_0 \sum_{n=0}^\infty \mathcal{J}_{2n+1}(a_\theta)\sin\left((2n+1)\omega\tau\right)\right\} \\
&\cong A\cos\theta_0 \mathcal{J}_0(a_\theta)t - \frac{4A\sin\theta_0}{\omega}\mathcal{J}_1(a_\theta)\sin^2(\omega t/2),
\end{aligned} \tag{37}$$

where $\mathcal{J}_n(x)$ is the $n$-th order Bessel function of the first kind. The approximation in the last line is valid for the present scenarios with $a_\theta \ll 1$.

## B. PROBING THE TOPOLOGY OF A TWO-LEVEL SYSTEM

### B.1 Initial state preparation and verification

In our experiments, the first step is to prepare the NV center spin into the ground state of the unperturbed Hamiltonian $\mathcal{H}(\theta_0, \varphi_0)$ as follows

$$\mathcal{H}(\theta_0, \varphi_0) = \frac{A}{2}\begin{pmatrix} \cos\theta_0 & \sin\theta_0 e^{-i\varphi_0} \\ \sin\theta_0 e^{i\varphi_0} & -\cos\theta_0 \end{pmatrix}. \tag{38}$$

which is a superposition state $|\psi(0)\rangle = \cos\frac{\theta}{2}|-1\rangle + e^{i\varphi_0}\sin\frac{\theta}{2}|0\rangle$, see Eq.(41), in which we denote $|\uparrow\rangle \equiv |-1\rangle$ and $|\downarrow\rangle \equiv |0\rangle$. Such an initial state can be prepared by applying a resonant microwave driving field as

$$\mathcal{H}_p = \frac{\omega_0}{2}\sigma_z + \Omega\sin(\omega_0 t + \varphi)\sigma_x, \tag{39}$$

for time duration $t = \theta_0/\Omega$, where $\Omega$ is the Rabi frequency. If the initial state is not in the eigenstate of the Hamiltonian $\mathcal{H}(\theta_0, \varphi_0)$, it would cause oscillation that may blur the coherent oscillation arising from parametric modulation. In order to verify that we have indeed prepared the NV center spin into the right initial state, we perform experiments by engineering a microwave driving field corresponding to the Hamiltonian in Eq.(52). If the initial state is the ground state of $\mathcal{H}(\theta_0, \varphi_0)$, no transition to the excited state will be observed. In the experiments, we carefully tune the microwave pulse duration for the initial state preparation so that no transition to the excited state under the Hamiltonian $\mathcal{H}(\theta_0, \varphi_0)$ occurs, as shown in Fig.1.

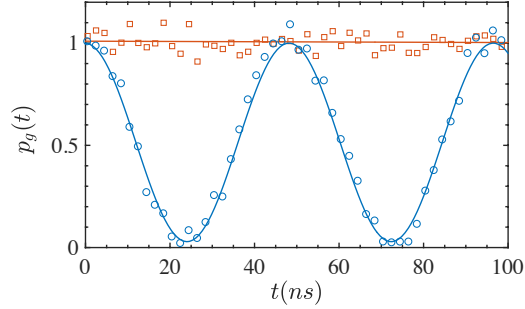

FIG. 1. Verification of initial state preparation. The state evolution from the initial state $|\psi(0)\rangle$, as governed by the Hamiltonian $\mathcal{H}(\theta_0, \varphi_0)$, is quantified by the fidelity $p_g(t) = |\langle\psi(t)|\psi(0)\rangle|^2$. The plot shows two initial states that are prepared by the microwave pulse $\mathcal{H}_i(t) = \Omega\sin(\omega_0 t + \varphi_0)\sigma_x$ with different time duration $t = 11.94$ ns (red, □) and $t = 0.04$ ns (blue, ○).

### B.2 Calibration of Rabi frequency and microwave amplitude

As we program parametric modulation in the waveform of AWG by specifying the microwave output amplitude, it is necessary to calibrate the relation between the microwave amplitude of AWG output and the Rabi frequency of the NV center spin. In

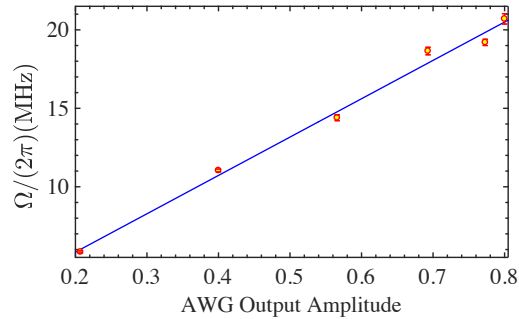

FIG. 2. Calibration of Rabi frequency and microwave amplitude. The measured Rabi frequency $\Omega$ of the NV center spin as a function of the output amplitude (in the unit of 500 mV [Vpp]) of arbitrary waveform generator while using the same microwave amplifying efficiency.

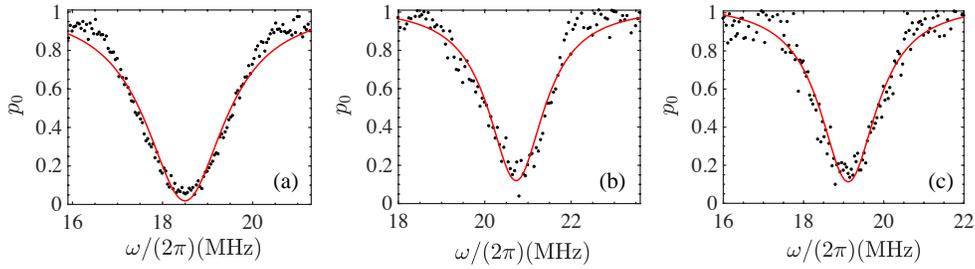

FIG. 3. Parametric modulation resonance measurement. The probability that the NV center spin remains in the ground state $|\psi(g)\rangle$ at time $T$ as a function of the modulation frequency $\omega$ of (a) linear parametric modulation with $a_\theta = 0$, $a_\varphi = 0.08$, and $(\theta_0, \varphi_0) = (\pi/2, 0)$, $T = 450$ ns; (b) linear parametric modulation with $a_\theta = a_\varphi = 0.1$, and $(\theta_0, \varphi_0) = (3\pi/4, 0)$, $T = 400$ ns; (c) elliptical parametric modulation with $a_\theta = a_\varphi = 0.1$, and $(\theta_0, \varphi_0) = (\pi/6, 0)$, $T = 450$ ns.

Fig.2, we show one example of such a calibration, which shows that the Rabi frequency of the NV center spin scales linear with the AWG output amplitude for the parameter regime in which our experiments are performed. The microwave calibration is done for all the measurements.

### B.3 Parametric modulation resonance measurement

In the parametric modulation resonance measurement experiments, we first prepare the NV center spin in the ground state $|\psi_0\rangle$ of the Hamiltonian $\mathcal{H}(\theta_0, \varphi_0)$ by a microwave pulse $\hat{n}(\varphi_0)|_{\theta_0}$. We then apply the engineered microwave driving field with different types of parametric modulation and fix the time duration $T$. By sweeping the parametric modulation frequency $\omega$, we measure the probability $p_0(T)$ that the NV center spin remains in the ground state $|\psi_0\rangle$. When the parametric modulation frequency $\omega$ matches the transition frequency, we observe a resonance signal in the probability $p_0(T)$. In the main text, as shown in Fig.1(e), we provide an example of parametric modulation resonance measurement. In Fig.3, we show the parametric modulation resonance measurement data for the other types of parametric modulation.

### B.4 Precise determination of parametric modulation resonance and oscillation frequency

After preparing the NV center spin in the ground state $|\psi_0\rangle$ of the Hamiltonian $\mathcal{H}(\theta_0, \varphi_0)$ by a microwave pulse $\hat{n}(\varphi_0)|_{\theta_0}$, we perform parametric modulation resonance measurement by sweeping the parametric modulation frequency $\omega$, and measure the probability $p_0(T)$ that the NV center spin remains in the ground state $|\psi_g\rangle$, from which we can roughly determine the resonant frequency. The more accurate we determine the parametric modulation resonant frequency, the more precise we can extract the information on quantum geometry. Therefore, in the experiments, we optimise the parametric modulation frequency to achieve the best possible oscillation contrast in order to improve the accuracy in determining the Rabi frequency. Fig.4 shows coherent transition from the ground state to the excited state on parametric modulation with slightly different modulation frequency. It can be seen that when the resonant condition is better matched, the parametric modulation induced coherent oscillation demonstrates

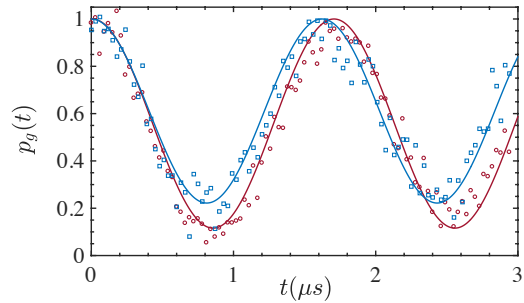

FIG. 4. Identification of exact resonance frequency. Coherent oscillation under elliptical parametric modulation with slightly different modulation frequency $\omega = (2\pi)20.98$ MHz (red, ○) and $\omega = (2\pi)20.82$ MHz (blue, □). The other parameters are: $\theta_0 = pi/4$, $\varphi = 0$, $a_\theta = 0.1$ and $a_\varphi = 0.1$.

a higher contrast. We observe that the Rabi frequencies depend on the modulation amplitudes, but also, on the form of the parametric modulation, which is shown to represent direct signatures of the system's quantum geometry.

### B.5 Detecting topological transition from measurement of quantum geometry

The topological transition can be detected in our experiment by extending the Hamiltonian into Eq.(12) with a tunable constant $r$ that can be realised with an additional frequency detuning. For different value of $r$, the QGT of the Hamiltonian Eq.(12) can be calculated analytically as

$$g_{\theta\theta} = \frac{(1 + r\cos\theta)^2}{4(1 + r^2 + 2r\cos\theta)^2}, \tag{40}$$

$$g_{\varphi\varphi} = \frac{\sin^2\theta}{4(1 + r^2 + 2r\cos\theta)}, \tag{41}$$

$$g_{\theta\varphi} = 0, \tag{42}$$

$$\mathcal{F}_{\theta\varphi} = \frac{\sin\theta(1 + r\cos\theta)}{2(1 + r^2 + 2r\cos\theta)^{3/2}}. \tag{43}$$

With a finite value of $r$, all elements of the QGT can also be measured directly in our experiment using the similar method as described in the main text. In this case, the frequency of parametric modulation would depend on $\theta$ because the eigenenergies of Hamiltonian Eq.(12) are

$$E_{\pm}(r, \theta, \varphi) = \pm\frac{A}{2}\sqrt{1 + r^2 + 2r\cos\theta}, \tag{44}$$

which change with the value of $\theta$. The NV center spin shall be initialised into the eigenstate

$$|\psi_g\rangle = \cos\frac{\theta'}{2}|-1\rangle + \sin\frac{\theta'}{2}e^{i\varphi}|0\rangle, \tag{45}$$

with

$$\theta' = \cos^{-1}\frac{\cos\theta + r}{\sqrt{1 + r^2 + 2r\cos\theta}} \tag{46}$$

which depends on the value of $r$.

When $r = 0$, we can look at the parameter space as a sphere $S^2$ embedded in $\mathbb{R}^3$, with a magnetic (Dirac) monopole at its origin. Changing the additional parameter $r$ moves the location of this monopole away from the origin, and the topology of this parameter space is characterized by the number of monopoles inside the sphere. The number of monopoles is counted by the Chern number, which is the integral of the Berry curvature over the sphere [2]

$$C = \frac{1}{2\pi}\int_{S^2}\mathcal{F}_{\theta\varphi}d\theta d\varphi. \tag{47}$$

For a finite value of $r$, the distribution of Berry curvature $\mathcal{F}_{\theta\varphi}$ is not symmetric on both sides of $\theta = \pi/2$. When $r$ approaches 1, the value of $\mathcal{F}_{\theta\varphi}$ has a singularity around $\theta = \pi$ which corresponds to a topological transition, where the monopole moves out of the sphere. In terms of the Chern number, the Chern number is $C = 1$ when $r < 1$, but $C = 0$ when $r > 1$.

### C. QUANTUM GEOMETRY MEASUREMENT OF AN INTERACTING TWO-QUBIT SYSTEM

#### C.1 Description and characterisation of the system

Coupling the NV center to a neighboring nuclear spin allows us to explore the quantum geometry and topology of an interacting two-qubit system. In our experiment, this two-qubit system consists of the NV center electron spin and a $^{13}$C nuclear spin located in the vicinity of the NV center. The nuclear spin of $^{14}$N is polarised, such that $m_{I^{(N)}} = +1$; in this setting, the total Hamiltonian of the NV electron spin (spin-1) and $^{13}$C nuclear spin (spin-$\frac{1}{2}$) can be written as

$$\mathcal{H} = D_{gs}S_z^2 + \gamma_e B_\parallel S_z + \gamma_n B_\parallel I_z + A_z S_z \otimes I_z + A_x S_z \otimes I_x, \tag{48}$$

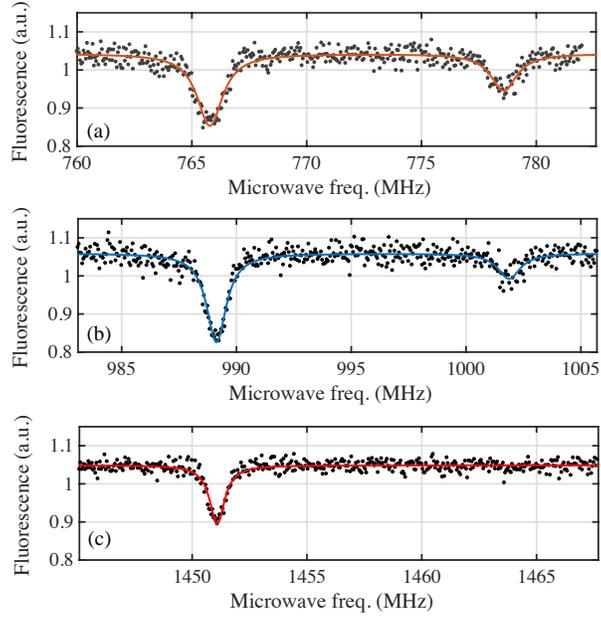

FIG. 5. Characterisation of spin-spin interaction. Pulsed optically detected magnetic resonance (pulsed ODMR) measurement with three different magnetic fields applying along the NV axis: $B_\parallel$ = 749.32 Gauss **(a)**, 669.64 Gauss **(b)**, 504.83 Gauss **(c)**. The two resonance dips corresponds to the transition frequencies $\omega_1$ and $\omega_2$ as in Eq.(49-50).

where $S$ is a spin-1 operator and where $I$ is the spin-$\frac{1}{2}$ operator; $D_{gs}$ = 2.87GHz is the zero-field splitting, $\gamma_e$ = 2.8MHz/G and $\gamma_n$ = 1.07kHz/G is the electronic spin and nuclear spin gyromagnetic ratio, respectively.

We perform pulsed optically detected magnetic resonance (pulsed ODMR) measurement (see Fig.5) in order to determine the coupling strength ($A_x$ and $A_z$) between the NV center electron spin and the $^{13}$C nuclear spin. Following the Hamiltonian in Eq.(48), the transition frequencies are given by [3]

$$\omega_1 = \omega_s - \frac{1}{2}\sqrt{A_x^2 + (A_z - \gamma_c B_\parallel)^2} - \frac{1}{2}\gamma_c B_\parallel, \quad (49)$$

$$\omega_2 = \omega_s + \frac{1}{2}\sqrt{A_x^2 + (A_z - \gamma_c B_\parallel)^2} + \frac{1}{2}\gamma_c B_\parallel, \quad (50)$$

where $\omega_s = D_{gs} - \gamma_e B_\parallel - A_N^{hs}$, $\gamma_e$ and $\gamma_c$ represent the gyromagnetic ratio of the NV center electron spin and $^{13}$C nuclear spin, respectively; $A_N^{hs} = -2.16$ MHz is the energy shift arising from the $^{14}$N nuclear spin projection $m_{I(N)} = +1$ associated with the NV center. The results for pulsed ODMR, as measured when applying three different magnetic fields along the NV axis, are shown in Fig.5. When combined with Eqs.(49)-(50), the measured transition frequencies $\omega_1$ and $\omega_2$ allow us to estimate the strength of the spin-spin interaction, and we obtain

$$A_x \approx 2.79\text{MHz} \qquad A_z \approx 11.832\text{MHz}. \quad (51)$$

We also point out that the pulsed ODMR measurement obtained by using a magnetic field of $B_\parallel$ = 505.13 Gauss along the NV axis [Fig.5(c)] only exhibits a single resonance; this result reflects the fact that the $^{13}$C nuclear spin is effectively polarised at the excited-state level anti-crossing (ESLAC) point; see Ref. [4].

### C.2 Details of topological property

In our experiment, we choose the NV center electronic spin manifold spanned by the spin sublevels $m_s = -1$ and $m_s = 0$ to encode the first qubit $|0\rangle \equiv |m_s = -1\rangle$, $|1\rangle \equiv |m_s = 0\rangle$. The nuclear spin of $^{13}$C encodes the second qubit $|0\rangle \equiv \left|+\frac{1}{2}\right\rangle$, $|1\rangle \equiv \left|-\frac{1}{2}\right\rangle$. Based on this configuration, we write the Hamiltonian describing these two coupled qubits [Eq. (48)] as

$$\mathcal{H} = \frac{\omega_0}{2}\sigma_z + \left(\frac{\gamma_n B_\parallel}{2} - \frac{A_z}{4}\right)\tau_z - \frac{A_x}{4}\tau_x - \frac{A_z}{4}\sigma_z \otimes \tau_z - \frac{A_x}{4}\sigma_z \otimes \tau_x, \quad (52)$$

where $\sigma$ and $\tau$ are the Pauli operators associated with the first and second qubit; $\omega_0 = D_{gs} - \gamma_e B$ is the energy splitting between the states $m_s = 0$ and $m_s = -1$ of the NV center electron spin, which is controlled through the magnetic field $B_\parallel$ applied along the NV axis. Applying an additional microwave field

$$\mathcal{H}_{\text{mw}} = \Omega_{\text{mw}} \sin\theta(t) \cos\left(\omega_0 t - 2\Omega_{\text{mw}} \int_0^t \cos\theta(\tau)d\tau + \varphi\right)\sigma_x, \tag{53}$$

one is able to rotate the electronic spin into an arbitrary direction; this protocol realizes the effective Hamiltonian in Eq.(13) of the main text. The Hamiltonian can be expressed in matrix form as

$$\mathcal{H}_{\text{rot}}(\theta, \varphi) = \frac{1}{2}\begin{pmatrix} \Omega_{\text{mw}}\cos\theta + \gamma_n B_\parallel - A_z & -A_x & \Omega_{\text{mw}}\sin\theta e^{-i\varphi} & 0 \\ -A_x & \Omega_{\text{mw}}\cos\theta - \gamma_n B_\parallel + A_z & 0 & \Omega_{\text{mw}}\sin\theta e^{-i\varphi} \\ \Omega_{\text{mw}}\sin\theta e^{i\varphi} & 0 & -\Omega_{\text{mw}}\cos\theta + \gamma_n B_\parallel & 0 \\ 0 & \Omega_{\text{mw}}\sin\theta e^{i\varphi} & 0 & -\Omega_{\text{mw}}\cos\theta - \gamma_n B_\parallel \end{pmatrix}, \tag{54}$$

where we used the basis

$$|0\rangle|0\rangle = |-1\rangle_e |+1/2\rangle_n, \, |1\rangle|0\rangle = |0\rangle_e |+1/2\rangle_n, \tag{55}$$
$$|0\rangle|1\rangle = |-1\rangle_e |-1/2\rangle_n, \, |1\rangle|1\rangle = |0\rangle_e |-1/2\rangle_n. \tag{56}$$

Henceforth, we denote the eigenstates of the Hamiltonian in Eq. (54) as $|\Psi_1\rangle$, $|\Psi_2\rangle$, $|\Psi_3\rangle$, $|\Psi_4\rangle$, according to their ordered eigenenergies $\epsilon_1 < \epsilon_2 < \epsilon_3 < \epsilon_4$.

Importantly, we find that the topological properties of the NV center are enriched by the interaction with the neighboring nuclear spin $^{13}$C. Indeed, in the single-qubit scenario, the Chern number of the ground state remains $C = 1$ for all values of the driving parameter $\Omega_{\text{mw}}$. In contrast, in the interacting-qubit setting, the spin-spin interaction favors zero Chern numbers for all eigenstates; hence, tuning the drive parameter $\Omega_{\text{mw}}$ allows one to drive topological phase transitions in this case. Specifically, the Chern number of the eigenstates $|\Psi_1\rangle$ and $|\Psi_3\rangle$ change from $C = 0$ to $C = 1$ after passing through the phase boundary corresponding to the critical value $\Omega_{\text{mw}}^{(c_1)} = \frac{1}{2}\left[-\gamma_n B_\parallel + \sqrt{(\gamma_n B_\parallel - A_z)^2 + A_x^2}\right]$. Similarly, the Chern number of the eigenstates $|\Psi_2\rangle$ and $|\Psi_4\rangle$ change from $C = 0$ to $C = -1$. The significance of the QGT in interacting systems is highlighted by its connection with the quantum Fisher information, which characterizes the statistical distinguishability between eigenstates [5]. In particular,

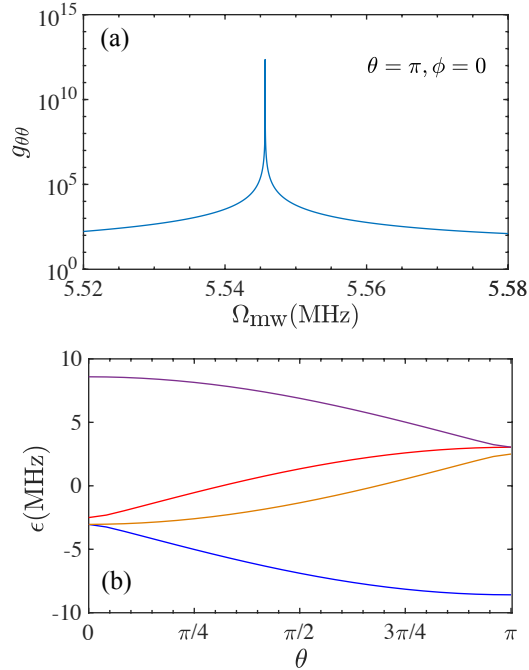

FIG. 6. (a) The Fubini-Study metric $g_{\theta\theta}$ (associated with the eigenstate $|\Psi_3\rangle$) as a function of $\Omega_{\text{mw}}$ close to the critical value $\Omega_{\text{mw}}^{(c_1)}$. (b) The eigenvalues as a function of the parameter $\theta$ for $\Omega_{\text{mw}} = 5.5456$ MHz close to the critical value.

the singular behaviour of the QGT is expected to be associated with quantum critical phenomena [5]; we remind that the quantum metric is related to the generalized susceptibility, and hence to quantum fluctuations via the fluctuation-dissipation theorem [6, 7]. In Fig. 6(a), we plot the Fubini-Study metric $g_{\theta\theta}$ (associated with the eigenstate $|\Psi_3\rangle$), which clearly shows a singular behaviour. The singularity arises due the nearly degenerate eigenstates around the critical point, see Fig.6(b).

We also notice that the system features a second phase boundary at the critical value

$$\Omega_{\text{mw}}^{(c_2)} = \frac{1}{2}\left[\gamma_n B_\parallel + \sqrt{(\gamma_n B_\parallel - A_z)^2 + A_x^2}\right], \tag{57}$$

where the eigenstates $|\Psi_2\rangle$ and $|\Psi_3\rangle$ undergo a topological phase transition associated with an exchange of the Chern number ($C = -1 \leftrightarrow C = 1$); we note that the topological properties of the eigenstates $|\Psi_1\rangle$ and $|\Psi_4\rangle$ are not modified through this process. We point out that in the regime $\Omega_{\text{mw}} \gg \Omega_{\text{mw}}^{(c_1)}$, the spin-spin interaction becomes less significant and we thus recover the topological properties of the two-level system, as we discussed in the case of two-level system.

### C.3 Extracting the QGT elements for interacting qubits

#### 1. Initial state preparation and verification

For arbitrary values of the parameters $(\theta, \phi)$, the eigenstates of the interacting Hamiltonian in Eq.(12) are superposition states, which can be written in the form

$$\begin{aligned}|\Psi\rangle &= \cos\vartheta\left(\cos\alpha_0 |-1\rangle_e + \sin\alpha_0 e^{i\beta_0} |0\rangle_e\right) \otimes |+1/2\rangle_c \\ &+ \sin\vartheta e^{i\eta}\left(\cos\alpha_1 |-1\rangle_e + \sin\alpha_1 e^{i\beta_1} |0\rangle_e\right) \otimes |-1/2\rangle_c.\end{aligned} \tag{58}$$

In our experiment, and without loss of generality, we have decided to measure the elements of the quantum geometry tensor associated with the eigenstate $|\Psi_3\rangle$. In fact, the initial state of the interacting spin system turns out to have a relatively high fidelity with the eigenstate $|\Psi_3\rangle$ when applying a magnetic field $B \sim 510$ Gauss (ESLAC) along the NV axis [see Fig.5(c)] together with a 532 nm polarizing laser pulse.

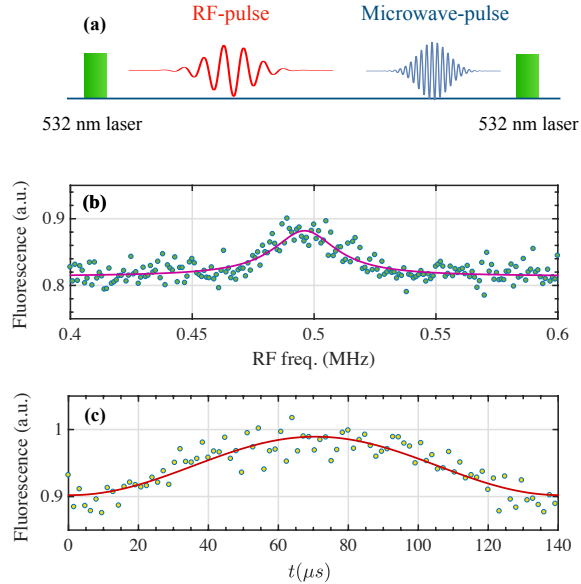

FIG. 7. Coherent manipulation of the $^{13}$C nuclear spin. (a) Experimental sequence to determine the energy splitting of the $^{13}$C nuclear spin and to observe Rabi oscillation: The first 532 nm laser pulse initialises the system and is followed by a RF-pulse acting on the $^{13}$C nuclear spin; the subsequent selective microwave-pulse flips the NV center electron spin state conditioning on the $^{13}$C nuclear spin state $|+1/2\rangle_c$ and the readout 532 nm laser pulse provides the information on the state population of the $^{13}$C nuclear spin. (b) The resonance when sweeping the frequency of the RF-pulse indicates that the energy splitting of the $^{13}$C nuclear spin (when the NV center spin is in the $m_s = 0$ state) is $\omega_c = 0.4961 \pm 0.0017$ MHz. (c) Rabi oscillation of the $^{13}$C nuclear spin.

In order to achieve the high fidelity of the initial state preparation, one can first apply a radio-frequency pulse and prepare the $^{13}$C nuclear spin into the following state

$$|\psi\rangle_c = \cos\vartheta\,|+1/2\rangle_c + \sin\vartheta e^{i\eta}\,|-1/2\rangle_c. \tag{59}$$

When the NV center electron spin is in the $m_s = 0$ state, the energy splitting of the $^{13}$C nuclear spin is $\omega_c = \gamma_c B_\parallel$. As shown in Fig.7(a), we sweep the frequency of the radio-frequency pulse (while fixing the pulse length), and the subsequent microwave pulse selectively flips the NV center electron spin state conditioning on the $^{13}$C nuclear spin state $|+1/2\rangle_c$. The resonance observed in Fig.7(b) indicates that the energy splitting of the $^{13}$C nuclear spin is $\omega_c = 0.4961 \pm 0.0017$ MHz. By applying a radio-frequency field at such a resonant frequency, we then measure Rabi oscillations of the $^{13}$C nuclear spin, see Fig.7(c). Subsequently, two selective microwave pulses that excite the electronic transition $|-1\rangle \leftrightarrow |0\rangle$ conditioning on the $^{13}$C nuclear spin state $|+1/2\rangle_c$ and $|-1/2\rangle_c$, respectively, would prepare the whole system into the target eigenstate $|\Psi\rangle$ in the form given in Eq. (58). The protocol for the verification of the system's initial state is similar to that described in the above section B.1.

### 2. Rabi oscillations induced by parametric modulation

Our experiment aims to extract all the elements of the quantum geometric tensor (QGT) associated with an eigenstate of the interacting two-qubit setting; as explained above, we take this eigenstate to be $|\Psi_3\rangle$. To achieve such a goal, we engineer a microwave driving field associated with a specific parametric modulation, and we extract the relevant coupling matrix elements (generalized Rabi frequencies) entering the QGT through Rabi-oscillation measurements. A priori, the Rabi oscillations induced by parametric modulations involve all the eigenstates of the interacting-spin system: they take place between the state of reference $|\Psi_3\rangle$ and all the other three eigenstates $|\Psi_1\rangle$, $|\Psi_2\rangle$ and $|\Psi_4\rangle$. However, we find that the Rabi frequency related to the transition between $|\Psi_3\rangle$ and $|\Psi_1\rangle$ is an order of magnitude larger as compared to all other transitions (to $|\Psi_2\rangle$, $|\Psi_4\rangle$), when using the same experimental parameters. In this sense, the elements of the QGT related to the eigenstate $|\Psi_3\rangle$ are dominated by the generalized Rabi frequencies associated with the coherent transition between $|\Psi_3\rangle$ and $|\Psi_1\rangle$; we note that the contribution to the QGT scales as $\sim \Omega^2$ in terms of the generalized Rabi frequency $\Omega$, see e.g. Eq.(24). In Fig.8, we plot the measured Rabi frequency $\Omega$ related to the coherent transition between the states $|\Psi_3\rangle$ and $|\Psi_1\rangle$, under linear and elliptical parametric modulations. These measurements allowed us to extract the complete quantum geometry tensor, which is displayed in Fig.5 (in the main text) together with theoretical predictions based on the Hamiltonian in Eq.(13).

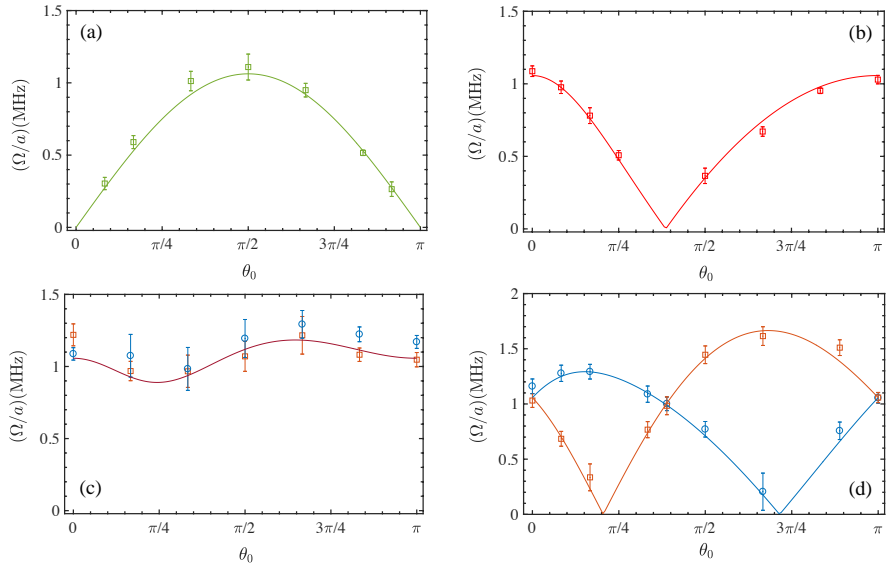

FIG. 8. Rabi frequency $\Omega$ of coherent transitions induced by parametric modulations, as a function of $\theta$ for four types of parametric modulations: (a-b) Linear parametric modulation with $a_\phi = a$ and $a_\theta = 0$ (a); $a_\phi = 0$ and $a_\theta = a$ (b). (c) Linear parametric modulation with $a_\theta = a$ and $a_\phi = a$ (brown, □); $a_\theta = a$ and $a_\phi = -a$ (blue, ○) (c). (d) Elliptical modulation with $a_\theta = a$ and $a_\phi = a$ (blue, ○); $a_\theta = a$ and $a_\phi = -a$ (brown, □). The experiment parameters are: $\Omega_{\text{mw}} = 2.13$ MHz. The curves represent the theoretical predictions.




\begin{thebibliography}{10}


\bibitem{01} T. W. B. Kibble, {\it Geometrization of quantum mechanics}, \href{https://link.springer.com/article/10.1007/BF01225149}{Comm. Math. Phys. \textbf{65}, 189-201 (1979)}.

\bibitem{02} J. Provost, G. Vallee, {\it Riemannian structure on mainfolds of quantum states}, \href{https://link.springer.com/article/10.1007/BF02193559}{Commun. Math. Phys. \textbf{76}, 289-301 (1980)}.

\bibitem{03} J. Anandan, Y. Aharonov, {\it Geometry of quantum evolution}, \href{https://journals.aps.org/prl/abstract/10.1103/PhysRevLett.65.1697}{Phys. Rev. Lett. \textbf{65}, 1697 (1990)}.

\bibitem{04} D. C. Brody, L. P. Hughston, {\it Geometric quantum mechanics}, \href{https://www.sciencedirect.com/science/article/pii/S0393044000000528}{J. Geom. Phys. \textbf{38}, 19-53 (2001)}.

\bibitem{05} M. Kolodrubetz, D. Sels, P. Mehta, A. Polkovnikov, {\it Geometry and non-adiabatic response in quantum and classical system}, \href{https://www.sciencedirect.com/science/article/pii/S0370157317301989}{Phys. Rep. \textbf{697}, 1-87 (2017)}.

\bibitem{06} B. Simon, {\it Holonomy, the quantum adiabatic theorem, and Berry's phase}, \href{https://journals.aps.org/prl/abstract/10.1103/PhysRevLett.51.2167}{Phys. Rev. Lett. \textbf{51}, 2167 (1983)}.

\bibitem{07} M. V. Berry, {\it Quantal phase factors accompanying adiabatic changes}, \href{https://royalsocietypublishing.org/doi/abs/10.1098/rspa.1984.0023}{Proc. R. Soc. London, Ser. A \textbf{392}, 45-57 (1984)}.

\bibitem{08} A. Bohm, A. Mostafazadeh, H. Koizumi, Q. Niu, J. Zwanziger, {\it The Geometric Phase in Quantum Systems: Foundations, Mathematical Concepts, and Applications in Molecular and Condensed Matter Physics}, \href{https://books.google.com/books?id=ydrzCAAAQBAJ}{(Springer Science \& Business Media 2013)}.

\bibitem{09} N. Nagaosa, J. Sinova, S. Onoda, A. H. MacDonald, N. P. Ong, {\it Anomalous Hall effect}, \href{https://journals.aps.org/rmp/abstract/10.1103/RevModPhys.82.1539}{Rev. Mod. Phys. \textbf{82}, 1539, (2010)}.

\bibitem{10} M. Z. Hasan, C. L. Kane, {\it Colloquium: Topological insulators}, \href{https://journals.aps.org/rmp/abstract/10.1103/RevModPhys.82.3045}{Rev. Mod. Phys. \textbf{82}, 3045 (2010)}.

\bibitem{11} I. Souza, T. Wilkens, R. M. Martin, {\it Polarization and localization in insulators: Generating function approach}, \href{https://journals.aps.org/prb/abstract/10.1103/PhysRevB.62.1666}{Phys. Rev. B \textbf{62}, 1666 (2000)}.

\bibitem{12} T. Ozawa and N. Goldman, {\it Probing localization and quantum geometry by spectroscopy}, \href{https://arxiv.org/abs/1904.11764}{arXiv:1904.11764}.

\bibitem{13} P. Zanardi, P. Giorda, M. Cozzini, {\it Information-theoretic differential geometry of quantum phase transitions}, \href{https://journals.aps.org/prl/abstract/10.1103/PhysRevLett.99.100603}{Phys. Rev. Lett. \textbf{99}, 100603 (2007)}.

\bibitem{14} V. V. Albert, B. Bradlyn, M. Fraas, L. Jiang, {\it Geometry and response of Lindbladians}, \href{https://journals.aps.org/prx/abstract/10.1103/PhysRevX.6.041031}{Phys. Rev. X \textbf{6}, 041031 (2016)}.

\bibitem{15} Y. Gao, S. A. Yang, Q. Niu, {\it Geometrical effects in orbital magnetic susceptibility}, \href{https://journals.aps.org/prb/abstract/10.1103/PhysRevB.91.214405}{Phys. Rev. B \textbf{91}, 214405 (2015)}.

\bibitem{16} F. Pi\'{e}chon, A. Raoux, J.-N. Fuchs, G. Montambaux, {\it Geometric orbital susceptibility: Quantum metric without Berry curvature}, \href{https://journals.aps.org/prb/abstract/10.1103/PhysRevB.94.134423}{Phys. Rev. B \textbf{94}, 134423 (2016)}.

\bibitem{17} O. Bleu, G. Malpuech, Y. Gao, and D. D. Solnyshkov, {\it Effective Theory of Nonadiabatic Quantum Evolution Based on the Quantum Geometric Tensor}, \href{https://journals.aps.org/prl/abstract/10.1103/PhysRevLett.121.020401}{Phys. Rev. Lett. \textbf{121}, 020401 (2018)}.

\bibitem{18} M. F. Lapa and T. L. Hughes, {\it Semiclassical wave packet dynamics in nonuniform electric fields}, \href{https://journals.aps.org/prb/abstract/10.1103/PhysRevB.99.121111}{Phys. Rev. B \textbf{99}, 121111(R) (2019)}.

\bibitem{19} A. Srivastava, A. Imamo\u{g}lu, {\it Signatures of Bloch-band geometry on excitons: Nonhydrogenic spectra in transition metal Dichalcogenides}, \href{https://journals.aps.org/prl/abstract/10.1103/PhysRevLett.115.166802}{Phys. Rev. Lett. \textbf{115}, 166802 (2015)}.

\bibitem{20} A. Julku, S. Peotta, T. I. Vanhala, D.-H. Kim, P. T\"{o}rm\"{a}, \href{https://journals.aps.org/prl/abstract/10.1103/PhysRevLett.117.045303}{{\it Geometric origin of superfluidity in the Lieb-Lattice flat band}, Phys. Rev. Lett. \textbf{117}, 045303 (2016)}.

\bibitem{21} R. Roy, {\it Band geometry of fractional topological insulators}, \href{https://journals.aps.org/prb/abstract/10.1103/PhysRevB.90.165139}{Phys. Rev. B \textbf{90}, 165139 (2014)}.

\bibitem{22} G. Palumbo, N. Goldman, {\it Revealing tensor monopoles through quantum-metric measurements}, \href{https://journals.aps.org/prl/abstract/10.1103/PhysRevLett.121.170401}{Phys. Rev. Lett. \textbf{121}, 170401 (2018)}.

\bibitem{23} P. Hauke, Markus. Heyl, L. Tagliacozzo and P. Zoller, {\it Measuring multipartite entanglement through dynamic susceptibilities}, \href{https://www.nature.com/articles/nphys3700}{Nature Physics {\bf 12}, 778 (2016)}.

\bibitem{24} T. Li, L. Duca, M. Reitter, F. Grusdt, E. Demler, M. Endres, M. Schleier-Smith, I. Bloch, U. Schneider, {\it Bloch state tomography using Wilson lines}, \href{https://science.sciencemag.org/content/352/6289/1094}{Science \textbf{352}, 1094 (2016)}.

\bibitem{25} L. Duca, T. Li, M. Reitter, I. Bloch, M. Schleier-Smith, U. Schneider, {\it An Aharonov-Bohm interferometer for determining Bloch band topology}, \href{https://science.sciencemag.org/content/347/6219/288}{Science \textbf{347}, 288 (2015)}.

\bibitem{26} N. Fl\"{a}schner, B. S. Rem, M. Tarnowski, D. Vogel, D.-S. L\"{u}hmann, K. Sengstock, C. Weitenberg, {\it Experimental reconstruction of the Berry curvature in a Floquet Bloch band}, \href{https://science.sciencemag.org/content/352/6289/1091}{Science \textbf{352}, 1091 (2016)}.

\bibitem{27} M. Wimmer, H. M. Price, I. Carusotto, U. Peschel, {\it Experimental measurement of the Berry curvature from anomalous transport} Nat. Phys. \textbf{13}, \href{https://www.nature.com/articles/nphys4050}{545-550 (2017)}.

\bibitem{28} H. B. Banks, Q. Wu, D. C. Valovcin, S. Mack, A. C. Gossard, L. Pfeiffer, R.-B. Liu, M. S. Sherwin, {\it Dynamical Birefringence: Electron-hole recollisions as probes of Berry curvature}, \href{https://journals.aps.org/prx/abstract/10.1103/PhysRevX.7.041042}{Phys. Rev. X \textbf{7}, 041042 (2017)}.

\bibitem{29} T. T. Luu, H. J. W\"{o}rner, {\it Measurement of the Berry curvature of solids using high-harmonic spectroscopy}, \href{https://www.nature.com/articles/s41467-018-03397-4}{Nat. Commun. \textbf{9}, 916 (2018)}.

\bibitem{30} N. Marzari, D. Vanderbilt, {\it Maximally localized generalized Wannier functions for composite energy bands}, \href{https://journals.aps.org/prb/abstract/10.1103/PhysRevB.56.12847}{Phys. Rev. B \textbf{56}, 12847 (1997)}.

\bibitem{31} L. Asteria, D. T. Tran, T. Ozawa, M. Tarnowski, B. S. Rem, N.Fl\"{a}schner, K. Sengstock, N. Goldman, C. Weitenberg, {\it Measuring quantized circular dichroism in ultracold topological matter}, \href{https://www.nature.com/articles/s41567-019-0417-8}{Nat. Phys. \textbf{15}, 449-454 (2019)}.

\bibitem{32} T. Ozawa and N. Goldman, {\it Extracting the quantum metric tensor through periodic driving}, \href{https://journals.aps.org/prb/abstract/10.1103/PhysRevB.97.201117}{Phys. Rev. B \textbf{97}, 201117(R) (2018)}.

\bibitem{33} T. Neupert, C. Chamon, and C. Mudry, {\it Measuring the quantum geometry of Bloch bands with current noise}, \href{https://journals.aps.org/prb/abstract/10.1103/PhysRevB.87.245103}{Phys. Rev. B \textbf{87}, 245103 (2013)}.

\bibitem{34} M. Kolodrubetz, V. Gritsev, and A. Polkovnikov, {\it Classifying and measuring geometry of a quantum ground state manifold}, \href{https://journals.aps.org/prb/abstract/10.1103/PhysRevB.88.064304}{Phys. Rev. B \textbf{88}, 064304 (2013)}.

\bibitem{35} O. Bleu, D. D. Solnyshkov, and G. Malpuech, {\it Measuring the quantum geometric tensor in two-dimensional photonic and exciton-polariton systems}, \href{https://journals.aps.org/prb/abstract/10.1103/PhysRevB.97.195422}{Phys. Rev. B \textbf{97}, 195422 (2018)}. 

\bibitem{36} P. Roushan, C. Neill, Y. Chen, M. Kolodrubetz, C. Quintana, N. Leung, M. Fang, R. Barends, B. Campbell, Z. Chen, B. Chiaro, A. Dunsworth, E. Jeffrey, J. Kelly, A. Megrant, J. Mutus, P. J. J. O'Malley, D. Sank, A. Vainsencher, J. Wenner, T. White, A. Polkovnikov, A. N. Cleland, J. M. Martinis, {\it Observation of topological transitions in interacting quantum circuits}, \href{https://www.nature.com/articles/nature13891}{Nature \textbf{515}, 241-244 (2014)}.

\bibitem{37} V. Gritsev, A. Polkovnikov, {\it Dynamical quantum Hall effect in the parameter space}, \href{https://www.pnas.org/content/109/17/6457.short}{Proc. Natl. Acad. Sci. USA \textbf{109}, 6457-6462 (2012)}.

\bibitem{38} M. D. Schroer, M. H. Kolodrubetz, W. F. Kindel, M. Sandberg, J. Gao, M. R. Vissers, D. P. Pappas, A. Polkovnikov, K. W. Lehnert, {\it Measuring a topological transition in an artificial spin-1/2 system}, \href{https://journals.aps.org/prl/abstract/10.1103/PhysRevLett.113.050402}{Phys. Rev. Lett. \textbf{113}, 050402 (2014)}.

\bibitem{39} D. T. Tran, N. R. Cooper, N. Goldman, {\it Quantized Rabi oscillations and circular dichroism in quantum Hall systems}, \href{https://journals.aps.org/pra/abstract/10.1103/PhysRevA.97.061602}{Phys. Rev. A \textbf{97}, 061602(R) (2018)}.

\bibitem{40} F. de Juan, A. G. Grushin, T. Morimoto, J. E. Moore, {\it Quantized circular photogalvanic effect in Weyl semimetals}, \href{https://www.nature.com/articles/ncomms15995}{Nat. Commun. \textbf{8}, 15995 (2017)}.

\bibitem{41} D. T. Tran, A. Dauphin, A. G. Grushin, P. Zoller, N. Goldman, {\it Probing topology by ``heating'': Quantized circular dichroism in ultracold atoms}, \href{https://advances.sciencemag.org/content/3/8/e1701207.short}{Sci. Adv. \textbf{3(8)}, e1701207 (2017)}.

\bibitem{42} C. Wang, P. Zhang, X. Chen, J. Yu, H. Zhai, {\it Scheme to measure the topological number of a Chern insulator from quench dynamics}, \href{https://journals.aps.org/prl/abstract/10.1103/PhysRevLett.118.185701}{Phys. Rev. Lett. \textbf{118}, 185701 (2017)}.

\bibitem{43} N. Fl\"{a}schner, D. Vogel, M. Tarnowski, B. S. Rem, D.-S. L\"{u}hmann, M. Heyl, J. C. Budich, L. Mathey, K. Sengstock, C. Weitenberg, {\it Observation of dynamical vortices after quenches in a system with topology}, \href{https://www.nature.com/articles/s41567-017-0013-8}{Nat. Phys. \textbf{14}, 265-268 (2018)}.

\bibitem{44} M. Tarnowski, F. N. \"{U}nal, N. Fl\"{a}schner, B. S. Rem, A. Eckardt, K. Sengstock, C. Weitenberg, {\it Measuring topology from dynamics by obtaining the Chern number from a linking number}, \href{https://www.nature.com/articles/s41467-019-09668-y}{Nat. Commun. \textbf{10}, 1728 (2019)}.

\bibitem{45} Materials and methods are available as supplementary material.

\bibitem{46} C. Repellin and N. Goldman, {\it Detecting fractional Chern insulators through circular dichroism}, \href{https://journals.aps.org/prl/abstract/10.1103/PhysRevLett.122.166801}{Phys. Rev. Lett. \textbf{122}, 166801 (2019)}.

\bibitem{47} M. Yu, P. Yang, M. Gong, Q. Cao, Q. Lu, H. Liu, M. B. Plenio, F. Jelezko, T. Ozawa, N. Goldman, S. Zhang and J. Cai, {\it Experimental measurement of the complete quantum geometry of a solid-state spin system}, \href{https://arxiv.org/abs/1811.12840}{arXiv:1811.12840}.


\bibitem{48} A. Gianfrate, O. Bleu, L. Dominici, V. Ardizzone, M. De Giorgi, D. Ballarini, K. West, L. N. Pfeiffer, D. D. Solnyshkov, D. Sanvitto, G. Malpuech, {\it Direct measurement of the quantum geometric tensor in a two-dimensional continuous medium}, \href{https://arxiv.org/abs/1901.03219}{arXiv:1901.03219}.

\bibitem{49} X. Tan, D.-W. Zhang, Z. Yang, J. Chu, Y.-Q. Zhu, D. Li, X. Yang, S. Song, Z. Han, Z. Li, Y. Dong, H.-F. Yu, H. Yan, S.-L. Zhu, and Y. Yu, {\it Experimental Measurement of the Quantum Metric Tensor and Related Topological Phase Transition with a Superconducting Qubit}, \href{https://journals.aps.org/prl/abstract/10.1103/PhysRevLett.122.210401}{Phys. Rev. Lett. {\bf 122}, 210401 (2019)}.

\vspace{0.1in}

\end{thebibliography}
\end{document}